\documentclass[aps,pra,showpacs,superscriptaddress,preprint]{revtex4}
\usepackage{amssymb}
\usepackage{amsmath}
\usepackage{graphicx}

\catcode`ð=\active
 \defð{\u{g}}
 \catcode`Ð=\active
 \defÐ{\u{G}}
 \catcode`Ý=\active
\defÝ{\. I}
 \catcode`ö=\active
\defö{\"{o}}
 \catcode`Ö=\active
 \defÖ{\"O}
 \catcode`ü=\active
 \defü{\"{u}}
 \catcode`Ü=\active
 \defÜ{\"{U}}
 \catcode`Þ=\active
\defÞ{\c{S}}
 \catcode`þ=\active
 \defþ{\c{s}}
 \catcode`ý=\active
 \defý{{\i}}
 \catcode`ç=\active
\defç{\d{c}}
 \catcode`Ç=\active
\defÇ{\d{C}}

\input{tcilatex}

\begin{document}

\title{Approximations to the bound states of Dirac-Hulth$\mathbf{{\acute{e}}}
$n problem}
\author{Sameer M. Ikhdair}
\email[E-mail: ]{sikhdair@neu.edu.tr}
\affiliation{Physics Department, Near East University, Nicosia, North Cyprus, Turkey}
\author{Ramazan Sever}
\email[E-mail: ]{sever@metu.edu.tr}
\affiliation{Physics Department, Middle East Technical University, 06800, Ankara, Turkey}
\date{%
\today%
}

\begin{abstract}
The bound state (energy spectrum and two-spinor wave functions) solutions of
the Dirac equation with the Hulth$\mathbf{{\acute{e}}}$n potential for all
angular momenta based on the spin and pseudospin symmetry are obtained. The
parametric generalization of the Nikiforov-Uvarov method is used in the
calculations. The orbital dependency (spin-orbit and pseudospin-orbit
dependent coupling too singular $1/r^{2}$) of the Dirac equation are
included to the solution by introducing a more accurate approximation scheme
to deal with the centrifugal (pseudo-centrifugal) term. \ The approximation
is also made for the less singular $1/r$ orbital term in the Dirac equation
for a wider energy spectrum. The nonrelativistic limits are also obtained on
mapping of parameters.

\end{abstract}

\maketitle

\newpage

\section{Introduction}

The spin or pseudospin symmetry [1,2] investigated by the framework of the
Dirac equation is one of the most interesting phenomena in the relativistic
quantum mechanics to explain different aspects for nucleon spectrum in
nuclei. This is mainly studied for the existence of identical bands in
superdeformed nuclei in the framework of a Dirac hamiltonian with attractive
scalar $S(\vec{r})$ and repulsive vector $V(\vec{r})$ potentials [3]. The
pseudospin symmetry is based on the small energy difference between
single-nucleon doublets with different quantum numbers and the Hamiltonian
of nucleons moving in the relativistic mean field produced by the
interactions between nucleons. The relativistic dynamics are described by
using the Dirac equation only [4].

Ginocchio [5] found that the pseudospin symmetry concept in nuclei occurs
when $S(\vec{r})$ and $V(\vec{r})$ potentials are nearly equal to each other
in magnitude but opposite in sign, i.e., $S(\vec{r})\sim -V(\vec{r})$ and
hence their sum is a constant, \textit{i.e.}, $\Sigma (r)=V(\vec{r})+S(\vec{r%
})=C_{ps}$. A necessary condition for occurrence of the pseudospin symmetry
in nuclei is to consider the case $\Sigma (\vec{r})=0$ [5-7]. Further, Meng 
\textit{et al} [8] showed that the pseudospin symmetry is exact under the
condition of $d\Sigma (\vec{r})/dr=0$. Lisboa \textit{et al} [9] studied the
generalized harmonic oscillator for spin-$1/2$ particles under the condition 
$\Sigma (\vec{r})=0$ or $\Delta (\vec{r})=V(\vec{r})-S(\vec{r})=0$. The
Dirac equation has been solved numerically [10,11] and analytically
[4,12,13] for nucleons that are moving independently in the relativistic
mean field in the presence of the pseudospin symmetric scalar and vector
potentials. The exact analytical solutions of the Dirac equation gives the
bound-state energy spectra and spinor wave functions [14,15].

The aim of this paper is to present an analytical bound state solutions of
the Dirac equation for the Hulth$\mathbf{{\acute{e}}}$n potential in the
presence of the exact pseudospin (spin) symmetry using a new approximation
scheme to deal with the pseudo-centrifugal (centrifugal) potential term for $%
\widetilde{l}>0$ ($l>0$) case. To obtain a general solution for all values
of the pseudospin (spin) quantum numbers, the pseudospin (spin) symmetry and
orbital dependency, pseudospin-orbit (spin-orbit) dependent coupling are
included to the lower component of the Dirac equation as an integer quantum
number. This component has the structure of the Schr\"{o}dinger-like
equations with the pseudo-centrifugal (spin-centrifugal) kinetic energy term
and its solution is analyzed by using some algebraic methods and effective
approaches. For small values of the radial coordinate $r$, this effective
potential gives a centrifugal energy term in the first approximation. The
Dirac equation for the Hulth$\mathbf{{\acute{e}}}$n potential is arranged
under the condition of the exact pseudospin (spin) symmetry and it's
solution is obtained systematically by using the Nikiforov-Uvarov (NU)
method [16]. As an application of the Dirac-Hulth$\mathbf{{\acute{e}}}$n
problem with the pseudospin (spin) symmetry, the relativistic eigenvalue
spectrum for various degenerate states is presented for several
pseudo-orbital (spin-orbital) and pseudospin (spin) quantum numbers.

The paper is organized as follows. In Sec. 2, the outline of the parametric
generalization of the NU method is presented. Section 3 is devoted for the
relativistic quantum mechanics (Dirac and Klein-Gordon equations) and the
additional coupling of the space scalar and vector potentials to free
particle wave equations. In Sec. 4, the pseudospin and spin symmetry Dirac
equation in $3+1$ dimensions with $1/r^{2}$ coupling is solved for the Hulth$%
\mathbf{{\acute{e}}}$n potential using an improved approximation scheme to
deal with the too singular pseudo-centrifugal (centrifugal) kinetic energy
term $\kappa \left( \kappa \pm 1\right) /r^{2}$. The parameteric
generalization of the NU method is followed to obtain the energy eigenvalues
and the corresponding two-spinor wave functions. In Sec. 5, we solve the
Dirac equation with an accurate proper approximation made for the less
singular coupling $1/r$ vector potential to extend the validity of the
results to a wider range energy spectrum. Results and conclusions are
presented in Sec. 6.

\section{Method of Analysis}

The Schr\"{o}dinger-like equation including the centrifugal barrier and/or
the spin-orbit coupling term has not been solved exactly for the
exponential-type potentials such as Morse, Hulth$\mathbf{{\acute{e}}}$n,
Woods-Saxon, etc [4,12-15,17]. The exact solution of the Schr\"{o}dinger
equation for the exponential-type potentials has been obtained for $l=0$,
however, any $l$-state solutions have been given approximately by using some
analytical methods under a certain number of restrictions [4,18-20]. One of
the calculational tools utilized in these studies is the Nikiforov-Uvarov
(NU) method. This technique is based on solving the hypergeometric type
second-order differential equations by means of the special orthogonal
functions [21]. For a given potential, the Schr\"{o}dinger or Schr\"{o}%
dinger-like equations in spherical coordinates are reduced to the
second-order differential equation of hypergeometric type with an
appropriate coordinate transformation $r\rightarrow s$ and then they are
solved systematically to find the exact or particular solutions. The NU
method is briefly outlined here: 
\begin{equation}
\psi _{n}^{\prime \prime }(r)+\frac{\widetilde{\tau }(r)}{\sigma (r)}\psi
_{n}^{\prime }(r)+\frac{\widetilde{\sigma }(r)}{\sigma ^{2}(r)}\psi
_{n}(r)=0,
\end{equation}%
where $\sigma (r)$ and $\widetilde{\sigma }(r)$ are polynomials, at most, of
second-degree, and $\widetilde{\tau }(r)$ is a first-degree polynomial. In
order to find a particular solution for Eq.(1), let us decompose the wave
function $\psi _{n}(r)$ as follows:%
\begin{equation}
\psi _{n}(r)=\phi (r)y_{n}(r),
\end{equation}%
and use%
\begin{equation}
\left[ \sigma (r)\rho (r)\right] ^{\prime }=\tau (r)\rho (r),
\end{equation}%
to reduce Eq.(1) to the form%
\begin{equation}
\sigma (r)y_{n}^{\prime \prime }(r)+\tau (r)y_{n}^{\prime }(r)+\lambdabar
y_{n}(r)=0,
\end{equation}%
with%
\begin{equation}
\tau (r)=\widetilde{\tau }(r)+2\pi (r),\text{ }\tau ^{\prime }(r)<0,
\end{equation}%
where the prime denotes the differentiation with respect to $r.$ One is
looking for a family of eigenvalue solutions corresponding to%
\begin{equation}
\lambdabar =\lambdabar _{n}=-n\tau ^{\prime }(r)-\frac{1}{2}n\left(
n-1\right) \sigma ^{\prime \prime }(r),\ \ \ n=0,1,2,\cdots .
\end{equation}%
The $y_{n}(r)$ can be expressed in terms of the Rodrigues relation:%
\begin{equation}
y_{n}(r)=\frac{B_{n}}{\rho (r)}\frac{d^{n}}{dr^{n}}\left[ \sigma ^{n}(r)\rho
(r)\right] ,
\end{equation}%
where $B_{n}$ is the normalization constant and the weight function $\rho
(r) $ is the solution of the differential equation (3). The other part of
the wave function (2) must satisfy the following logarithmic equation%
\begin{equation}
\frac{\phi ^{\prime }(r)}{\phi (r)}=\frac{\pi (r)}{\sigma (r)}.
\end{equation}%
By defining 
\begin{equation}
k=\lambdabar -\pi ^{\prime }(r).
\end{equation}%
one obtains the polynomial

\begin{equation}
\pi (r)=\frac{1}{2}\left[ \sigma ^{\prime }(r)-\widetilde{\tau }(r)\right]
\pm \sqrt{\frac{1}{4}\left[ \sigma ^{\prime }(r)-\widetilde{\tau }(r)\right]
^{2}-\widetilde{\sigma }(r)+k\sigma (r)},
\end{equation}%
where $\pi (r)$ is a parameter at most of order one$.$ The expression under
the square root sign in the above equation can be arranged as a polynomial
of second order where its discriminant is zero. Hence, an equation for $k$
is being obtained. After solving such an equation, the $k$ values are
determined through the NU method. \ 

We may also derive an alternative parameteric generalization from the NU
method valid for most potential models under consideration. The first step
basically begins by writting the hypergeometric equation [21] in general
parametric form as 
\begin{equation}
\left[ r\left( c_{3}-c_{4}r\right) \right] ^{2}\psi _{n}^{\prime \prime }(r)+%
\left[ r\left( c_{3}-c_{4}r\right) \left( c_{1}-c_{2}r\right) \right] \psi
_{n}^{\prime }(r)+\left( -\xi _{1}r^{2}+\xi _{2}r-\xi _{3}\right) \psi
_{n}(r)=0,
\end{equation}%
with%
\begin{equation}
\widetilde{\tau }(r)=c_{1}-c_{2}r,\text{ }\sigma (r)=\left(
c_{3}-c_{4}r\right) r,\text{ }\widetilde{\sigma }(r)=-\xi _{1}r^{2}+\xi
_{2}r-\xi _{3},
\end{equation}%
where the coefficients $c_{i}$ ($i=1,2,3,4$) and the analytic expressions $%
\xi _{j}$ ($j=1,2,3$) have to be calculated for the potential model under
consideration. The second step demands comparing Eq.(11) with it's
counterpart Eq.(1) so that we can obtain the analytic NU polynomials, energy
equation, wave functions and the relevant coefficients expressed in general
parameteric form in Appendix A of Ref. [22].

\section{Scalar Potential Coupling in Relativistic Quantum Mechanics}

The Dirac equation and the Klein-Gordon (KG) equation are wave equation
mostly used in describing particle dynamics in relativistic quantum
mechanics. These two wave equations, for free particles, are constructed
using two objects: the four-vector linear momentum operator $P_{\mu }=i\hbar
\partial _{\mu }$ and the scalar rest mass $M,$ allow one to introduce
naturally two types of potential coupling. One is the gauge-invariant
coupling to the four-vector potential $\left\{ A_{\mu }\left( t,%
\overrightarrow{r}\right) \right\} _{\mu =0}^{3}$ which is introduced via
the minimal substitution $P_{\mu }\rightarrow P_{\mu }-gA_{\mu },$ where $g$
is a real coupling parameter. The other, is an additional coupling to the
space-time scalar potential $S(t,\overrightarrow{r})$ which is introduced by
the substitution $M\rightarrow M+S.$ The term \textquotedblleft
four-vector\textquotedblright\ and \textquotedblleft
scalar\textquotedblright\ refers to the corresponding unitary irreducible
representation of the Poincar$\mathbf{{\acute{e}}}$ space-time symmetry
group (the group of rotations and translations in ($3+1$)-dimensional
Minkowski space-time). Gauge invariance of the vector coupling allows for
the freedom to fix the gauge (eliminating the non physical gauge modes)
without altering the physical content of the problem. Many choose to
simplify the solution of the problem by taking the space component of the
vector potential to vanish (i.e., $\overrightarrow{A}$). One may write the
time-component of the four-vector potential as $gA_{0}=V(t,\overrightarrow{r}%
),$ then it ends up with two independent potential functions in the Dirac
and KG equations. These are the \textquotedblleft vector\textquotedblright\
potential $V$ and the \textquotedblleft scalar\textquotedblright\ potential $%
S$ [23,24]$.$

In the relativistic units, $\hbar =c=1,$ the free Dirac and KG equations are
written as 
\begin{subequations}
\begin{equation}
\left( i\gamma ^{\mu }\partial _{\mu }-M\right) \psi _{D}(t,\overrightarrow{r%
})=0,
\end{equation}%
\begin{equation}
(\partial ^{\mu }\partial _{\mu }+M^{2})\psi _{KG}(t,\overrightarrow{r})=0,
\end{equation}%
respectively. The convention of summing over repeated indices is used. For
particles of spin $1/2,\left\{ \gamma ^{\mu }\right\} $ are $4\times 4$
constant matrices with the following standard representation [23]: 
\end{subequations}
\begin{equation*}
\gamma ^{0}=\left( 
\begin{array}{cc}
I & 0 \\ 
0 & -I%
\end{array}%
\right) ,\text{ }\overrightarrow{\gamma }=\left( 
\begin{array}{cc}
0 & \vec{\sigma} \\ 
-\vec{\sigma} & 0%
\end{array}%
\right) ,
\end{equation*}%
where $I$ is the $2\times 2$ unit matrix and $\vec{\sigma}$ are the three $%
2\times 2$ hermitian Pauli spin matrices. The vector and scalar couplings
mentioned above introduce potential interactions by mapping the free Dirac
and KG equations as 
\begin{subequations}
\begin{equation}
\left\{ \gamma ^{0}\left[ i\frac{\partial }{\partial t}-V(t,\overrightarrow{r%
})\right] +i\overrightarrow{\gamma }\cdot \overrightarrow{\nabla }-\left[
M+S(t,\overrightarrow{r})\right] \right\} \psi _{D}(t,\overrightarrow{r})=0,
\end{equation}%
\begin{equation}
\left\{ -\left[ i\frac{\partial }{\partial t}-V(\overrightarrow{r})\right]
^{2}-\overrightarrow{\nabla }^{2}+\left[ M+S(\overrightarrow{r})\right]
^{2}\right\} \psi _{KG}(\overrightarrow{r})=0,
\end{equation}%
respectively. This type of coupling attracted a lot of attention in the
literature due to the resulting simplification in the solution of the
relativistic problem. The scalar-like potential coupling is added to the
scalar mass so that in case when $S(r)=+V(r),$ the Dirac equation could
always be reduced to a Schr\"{o}dinger-type second order differential
equation as we shall see in the next section. The nonrelativistic limit can
be obtained by taking $E_{n\kappa }-M\simeq E_{nl}$ and $E+M\simeq 2M,$
where $\left\vert E\right\vert \ll M.$ Hence, the positive energy solution
is $\left[ \frac{1}{2M}\overrightarrow{\nabla }^{2}-2V(\overrightarrow{r}%
)+E_{nl}\right] \psi (\overrightarrow{r})=0,$ with potential $2V,$
nonrelativistic energy $E_{nl}$ and $\psi (\overrightarrow{r})$ stands for
either $\psi _{D}^{+}(\overrightarrow{r})$ or $\psi _{KG}(\overrightarrow{r}%
).$ The negative energy solution corresponding to case $S=-V$ results in a
trivial non-interacting theory (free fields) with solution $\left[ \frac{1}{%
2M}\overrightarrow{\nabla }^{2}+E_{nl}\right] \psi (\overrightarrow{r})=0$
[24]$.$ The physical meaning of introducing the scalar-like potential into
Dirac and KG equations is making one to study the confinements in quarks
when taking the conditions $S=V,$ $S=-V$ and $S=\eta V$ such that $\eta \neq
\pm 1.$ The last case results in uneven contribution of the two potentials
[24]. For example, suppose $S=V$ is a potential which tends to an effective
positive infinite barrier at spatial infinity for the positive-energy
particles and will be responsible for the confinement solutions (bound
states or scattering states). However, there is an effective infinite well
for the negative-energy particles which cannot prevent the negative-energy
particles from going to infinity [25]. This provides real (bound state
solutions) for the positive-energy particles, but imaginary (scattering
state solutions) for the negative-energy particles. Similarly, when $S=V$
tends to negative infinity at spatial infinity, the positive-energy
particles are not confined due to the effective potential well [25].
Therefore, the $(1+3)$- and $(1+1)$-dimensional Dirac equation with $S=V,$
the confinement is impossible, i.e., there must be scattering states. If $S$
is stronger than $V,$ the confinement is permanent and if, on the contrary,
the $V$ is stronger, confinement is impossible due to the Klein paradox (see
Ref. [25] and references therein).

\section{Dirac Equation with Coupling to $r^{-2}$ Singular Orbital Term}

The Dirac equation of a single-nucleon of rest mass $M$ with spherically
symmetric coupling to an attractive scalar and repulsive vector $S(\vec{r})$
and $V(\vec{r})$ potentials has the following radial component 
\end{subequations}
\begin{equation}
\left[ \vec{\alpha}.c\vec{P}+\beta (Mc^{2}+S(\vec{r}))\right] \psi _{n\kappa
}(\vec{r})=\left[ E_{n\kappa }-V(\vec{r})\right] \psi _{n\kappa }(\vec{r}),
\end{equation}%
where 
\begin{equation}
\vec{P}=-i\hbar \vec{\nabla},~\vec{\alpha}=\left( 
\begin{array}{cc}
0 & \vec{\sigma} \\ 
\vec{\sigma} & 0%
\end{array}%
\right) ,~\beta =\left( 
\begin{array}{cc}
0 & I \\ 
-I & 0%
\end{array}%
\right) ,
\end{equation}%
where $\vec{P}$ is the three momentum operators, $\vec{\alpha}$ and $\beta $
are the usual $4\times 4$ Dirac matrices [23], $c$ is the speed of light in
vacuum and $\hbar $ is the Planck's constant divided by $2\pi $. Further, $%
E_{n\kappa }$ denotes the relativistic energy eigenvalues of the Dirac
particle. For nuclei with spherical symmetry, $S(\vec{r})=S(r)$ and $V(\vec{r%
})=V(r)$, where $r$ is the magnitude of $\vec{r}$. Further, the spinor wave
functions is%
\begin{equation}
\psi _{n\kappa }(\vec{r})=\frac{1}{r}\left( 
\begin{array}{c}
F_{n\kappa }(r)\left[ Y_{_{l}}(\theta ,\phi )\chi _{\pm }\right] _{m}^{(j)}
\\ 
iG_{n_{r}\kappa }(r)\left[ Y_{_{\widetilde{_{l}}}}(\theta ,\phi )\chi _{\pm }%
\right] _{m}^{(j)}%
\end{array}%
\right) ,
\end{equation}%
where $Y_{l}(\theta ,\phi )$ ($Y_{\widetilde{l}}(\theta ,\phi )$) and $\chi
_{\pm }$ are the spin (pseudospin) spherical harmonic and spin wave function
which are coupled to angular momentum $j$ with projection $m$, respectively. 
$F_{n\kappa }(r)$ and $G_{n\kappa }(r)$ are the radial wave functions for
the upper and lower components, respectively. The label $\kappa $ has two
explanations; the aligned spin $j=l+1/2$ ($s_{1/2},p_{3/2},etc.$) is valid
for the case of $\kappa =-(j+1/2)$ and then $\widetilde{l}=l+1$, while the
unaligned spin $j=l-1/2$ ($p_{1/2},d_{3/2},etc.$) is valid for the case of $%
\kappa =(j+1/2)$ and then $\widetilde{l}=l-1$. Thus, the quantum number $%
\kappa $ and the radial quantum number $n$ are sufficient to label the Dirac
eigenstates. The Dirac equation (15) can be reduced to the following two
coupled ordinary differential equations (in the relativistic units, $\hbar
=c=1$): 
\begin{equation}
\left( \frac{d}{dr}+\frac{\kappa }{r}\right) F_{n\kappa }(r)=(M+E_{n\kappa
}-\Delta (r))G_{n\kappa }(r),
\end{equation}%
\begin{equation}
\left( \frac{d}{dr}-\frac{\kappa }{r}\right) G_{n\kappa }(r)=(M-E_{n\kappa
}+\Sigma (r))F_{n\kappa }(r),
\end{equation}%
where $\Delta (r)=V(r)-S(r)$ and $\Sigma (r)=V(r)+S(r)$ are the difference
and the sum potentials, respectively. Solving Eqs.(18) and (19) leads to a
second order Schr\"{o}dinger-like differential equation with coupling to $%
r^{-2}$ singular term and satisfying $G_{n\kappa }(r),$%
\begin{equation}
\left( \frac{d^{2}}{dr^{2}}-\frac{\kappa (\kappa -1)}{r^{2}}-(M+E_{n\kappa
}-\Delta (r))(M-E_{n\kappa }+\Sigma (r))-\frac{\frac{d\Sigma (r)}{dr}\left( 
\frac{d}{dr}-\frac{\kappa }{r}\right) }{M-E_{n\kappa }+\Sigma (r)}\right)
G_{n\kappa }(r)=0,
\end{equation}%
where $E_{n\kappa }\neq +M$ when $\Sigma (r)=0$ (exact pseudospin symmetry).
Since $E_{n\kappa }=-M$ is an element of the negative energy spectrum of the
Dirac Hamiltonian, then this relation with the lowe spinor component is not
valid for the positive energy solution. Further, a similar equation
satisfying $F_{n\kappa }(r)$ can be obtained as 
\begin{equation}
\left( \frac{d^{2}}{dr^{2}}-\frac{\kappa (\kappa +1)}{r^{2}}-(M+E_{n\kappa
}-\Delta (r))(M-E_{n\kappa }+\Sigma (r))+\frac{\frac{d\Delta (r)}{dr}\left( 
\frac{d}{dr}+\frac{\kappa }{r}\right) }{M+E_{n\kappa }-\Delta (r)}\right)
F_{n\kappa }(r)=0,
\end{equation}%
where $E_{n\kappa }\neq -M$ when $\Delta (r)=0$ (exact spin symmetry). Since 
$E_{n\kappa }=+M$ is an element of the positive energy spectrum of the Dirac
Hamiltonian, then this relation with the upper spinor component is not valid
for the negative energy solution. The exact spin symmetry requires ($d\Delta
(r)/dr=0$, \textit{i.e.}, $\Delta (r)=C_{s}=$constant), Eq. (21) turns out
to be 
\begin{equation}
\left( \frac{d^{2}}{dr^{2}}-\frac{\kappa (\kappa +1)}{r^{2}}-(M+E_{n\kappa
}-C_{s})\Sigma (r)+E_{n\kappa }^{2}-M^{2}+C_{s}\left( M-E_{n\kappa }\right)
\right) F_{n\kappa }(r)=0,
\end{equation}%
where $\kappa =l$ and $\kappa =-(l+1)$ are valid for $\kappa >0$ and $\kappa
<0,$ respectively, and $\kappa (\kappa +1)/r^{2}$ is the spin-centrifugal
potential term. On the other hand, the exact pseudospin symmetry requires ($%
d\Sigma (r)/dr=0$, \textit{i.e.}, $\Sigma (r)=C_{ps}=$constant), Eq. (20) is
reduced to the form%
\begin{equation}
\left( \frac{d^{2}}{dr^{2}}-\frac{\kappa (\kappa -1)}{r^{2}}+(M-E_{n\kappa
}+C_{ps})\Delta (r)+E_{n\kappa }^{2}-M^{2}-C_{ps}\left( M+E_{n\kappa
}\right) \right) G_{n\kappa }(r)=0,
\end{equation}%
where $\kappa =\widetilde{l}+1$ and $\kappa =-\widetilde{l}$ are valid for $%
\kappa >0$ and $\kappa <0,$ respectively, and $\kappa (\kappa -1)/r^{2}$ is
the pseudo-centrifugal potential term. Therefore, the degenerate states come
into existence with the same $\widetilde{l}$ but different $\kappa $,
generating pseudospin symmetry. The components of the wave function are
required to satisfy the boundary conditions. That is, $F_{n\kappa
}(r)/r\rightarrow 0$ ($G_{n\kappa }(r)/r$ $\rightarrow 0)$ when $%
r\rightarrow \infty $ and $F_{n\kappa }(r)/r=0$ ($G_{n\kappa }(r)/r=0)$ at $%
r=0$ hold. Note that the analytic solutions of the above second order
differential equations require approximation to the orbital term $\kappa
(\kappa \pm 1)r^{-2}$ that results from reduction of the original Dirac
equation. For example, the orbital term $\sim r^{-2}$ has a more singularity
near $r=0.$

\subsection{ Pseudospin Symmetry Solution}

The solution of the Dirac equation (23) for the Hulth$\mathbf{{\acute{e}}}$n
potential demands that the potential $\Delta (r)$ is exponential in $r$ and
the pseudo-centrifugal term is quadratic in $1/r$. Hence, the difference
potential is taken as the Hulth$\mathbf{{\acute{e}}}$n potential [26]: 
\begin{equation}
\Delta (r)=-\Delta _{0}\frac{e^{-\delta r}}{1-e^{-\delta r}},\text{ }\Delta
_{0}=V_{0}=Ze^{2}\delta ,
\end{equation}%
where $\delta $ and $\Delta _{0}$ are the screening parameters to determine
the range and strength, respectively. Besides, $Ze$ is the charge of the
nucleon [27]. This potential has been studied by means of the algebraic
perturbation calculations based upon the dynamical group structure SO(2,1)
[28], the NU method [29], the supersymmetry and shape invariance [30], the
asymptotic iteration method [31] and the Biedenharn's approach for the
Dirac-Coulomb problem [32]. Equation (23) is analytically solvable only for $%
\widetilde{l}=0$ ($\kappa =1$). Therefore, in order to solve the Dirac
equation for any $\kappa $ or $\widetilde{l}$-state, we need to apply the
following shifted approximation scheme near the singularity (origin) to deal
with the more singular pseudo-centrifugal term, $r^{-2},$ for the case of $%
\widetilde{l}>0$ [20,29,33]%
\begin{equation*}
\frac{1}{r^{2}}\approx \delta ^{2}\left[ d_{0}+\frac{1}{e^{\delta r}-1}+%
\frac{1}{(e^{\delta r}-1)^{2}}\right]
\end{equation*}%
\begin{equation}
=\underset{\delta \rightarrow 0}{\lim }\delta ^{2}\left[ d_{0}+\frac{1}{%
\left( \delta r\right) ^{2}}-\frac{1}{12}+\frac{\left( \delta r\right) ^{2}}{%
240}-\frac{\left( \delta r\right) ^{4}}{6048}+\frac{\left( \delta r\right)
^{6}}{172800}+O\left( \left( \delta r\right) ^{8}\right) \right] ,
\end{equation}%
where the approximation constraint is $\delta \rightarrow 0$ or applying the
condition $\delta r\rightarrow 0.$ It should be noted that the physical
interpretaion in introducing the parameter $d_{0}$ to the traditional
approximation, $\delta ^{2}e^{\delta r}/(e^{\delta r}-1)^{2},$ is that when
performing the power series expansion and letting $\delta r\rightarrow 0,$
it gives $r^{-2}-1/12$ but not $r^{-2}$ as desired. $\ $Th$\func{erf}$ore,
we understand that traditional (conventional) approximation scheme suggested
by Greene and Aldrich [34] is shifted by a dimensionless constant $%
d_{0}=1/12 $ from the origin. Equation (25) is the correct ansatz to
substitute $r^{-2}$ (see [33] and the references therein)$.$ The above
approximation resembles $r^{-2}\approx r^{-2}+\delta ^{2}f(d_{0}),$ where $%
f(d_{0})=d_{0}-1/12.$ Furtehr, it is simply the addition of traditional
(usual) approximation plus a shifting term $\delta ^{2}d_{0}$, i.e., $%
r^{-2}\approx \delta \left[ W(r)+W^{2}(r)/\delta +\delta d_{0}\right] ,$
with $W(r)=\delta /(e^{\delta r}-1).$ It is apparent from the above
expansion that for small values of $\delta $, the dimensionless constant $%
d_{0}=1/12.$ However, the approximation model used in [4] is $r^{-2}=W^{2}(r)
$ (cf. Ref. [4] and the references therein). Figure 1a shows a plot of the
variation of the centrifugal orbital term $r^{-2}$ with respect to $r,$
where the screening parameter $\delta =0.1$ $fm^{-1}.$ We observe that the
improved approximation model (solid line) works well if compared with $r^{-2}
$ (dotted-solid line). The curves in Fig. 1b show that the approximation of $%
r^{-2}$ is independent of the value of $d_{0}.$ The traditional
approximation is plotted as a function for different values of $d_{0}.$ It
should be noted that old approximation in Ref. [4] has the ansatz $%
r^{-2}\approx W^{2}(r)$ (i.e., one term function in (25)).

Now, the substitution of Eq.(24) and Eq. (25) into Eq. (23) leads to%
\begin{equation}
\left[ \frac{d^{2}}{dr^{2}}-\kappa (\kappa -1)\delta ^{2}\left( d_{0}+\frac{%
e^{-\delta r}}{(1-e^{-\delta r})^{2}}\right) -\delta ^{2}\left( \nu _{1}^{2}%
\frac{e^{-\delta r}}{1-e^{-\delta r}}+\omega _{1}^{2}\right) \right]
G_{n_{r}\kappa }(r)=0,
\end{equation}%
with%
\begin{equation}
\nu _{1}^{2}=\frac{(M-E_{n\kappa }+C_{ps})\Delta _{0}}{\delta ^{2}},\text{ }%
\omega _{1}^{2}=\frac{M^{2}-E_{n\kappa }^{2}+C_{ps}\left( M+E_{n\kappa
}\right) }{\delta ^{2}}.
\end{equation}%
Further, defining%
\begin{equation*}
s=e^{-\delta r}\in \lbrack 0,+1],\text{ }A_{1}=\omega _{1}^{2}-\nu
_{1}^{2}+\kappa \left( \kappa -1\right) d_{0},
\end{equation*}%
\begin{equation}
\text{ }B_{1}=2\omega _{1}^{2}-\nu _{1}^{2}+\kappa \left( \kappa -1\right)
\left( 2d_{0}-1\right) ,\text{ }\epsilon _{n\kappa }^{2}=\omega
_{1}^{2}+\kappa \left( \kappa -1\right) d_{0},
\end{equation}%
recasts Eq. (26) into the simple form 
\begin{equation}
\left( \frac{d^{2}}{ds^{2}}+\frac{1-s}{s(1-s)}\frac{d}{ds}+\frac{%
-A_{1}s^{2}+B_{1}s-\epsilon _{n\kappa }^{2}}{s^{2}(1-s)^{2}}\right)
G_{n\kappa }(s)=0,\text{ }G_{n\kappa }(1)=G_{n\kappa }(0)=0,
\end{equation}%
which can be easily solved by means of the NU method or applying a short-cut
procedure given in Appendix A of Ref. [22].

The procedures begin by comparing Eq. (29) with Eq. (1) giving the
polynomials: 
\begin{equation}
\widetilde{\tau }(s)=1-s,~~~{\sigma }(s)=s(1-s),~~~\widetilde{\sigma }%
(s)=-A_{1}s^{2}+B_{1}s-\epsilon _{n\kappa }^{2},
\end{equation}%
and with the aid of Eqs.(12)-(14), we can obtain $c_{i}=1$ (for $i=1,2,3,4$%
), $\xi _{1}=A_{1},$ $\xi _{2}=B_{1}$ and $\xi _{3}=\epsilon _{n\kappa
}^{2}. $ In addition, the relations A1-A3 yield%
\begin{equation*}
c_{5}=0,\text{ }c_{6}=-\frac{1}{2},\text{ }c_{7}=\frac{1}{4}+A_{1},\text{ }%
c_{8}=-B_{1},\text{ }c_{9}=\epsilon _{n\kappa }^{2},
\end{equation*}%
\begin{equation}
\text{ }c_{10}=\frac{\left( 2\kappa -1\right) ^{2}}{4},\text{ }%
c_{11}=2\epsilon _{n\kappa },\text{ }c_{12}=2\kappa -1,\text{ }%
c_{13}=\epsilon _{n\kappa },\text{ }c_{14}=\kappa ,
\end{equation}%
and the relations A4-A6 give the essential NU polynomials: 
\begin{equation*}
\pi (s)=\epsilon _{n\kappa }-\left( \kappa +\epsilon _{n\kappa }\right) s,%
\text{ }k=-\nu _{1}^{2}-\kappa \left( \kappa -1\right) -\left( 2\kappa
-1\right) \epsilon _{n\kappa },
\end{equation*}%
\begin{equation}
\tau (s)=1+2\epsilon _{n\kappa }-2\left( \epsilon _{n\kappa }+\kappa +\frac{1%
}{2}\right) s,\text{ }\tau ^{\prime }(s)=-2\left( \epsilon _{n\kappa
}+\kappa +\frac{1}{2}\right) <0.
\end{equation}%
The eigenvalue equations (6) and (9) take the forms%
\begin{equation}
\lambdabar _{n}=n^{2}+2n\left( \epsilon _{n\kappa }+\kappa \right) \text{
and }\lambdabar =-\nu _{1}^{2}-\kappa (2\epsilon _{n\kappa }+\kappa ),
\end{equation}%
respectively. In setting $\lambdabar =\lambdabar _{n}$ or alternatively
using the relation A7, we obtain the eigenvalue equation being expressed in
terms of $E_{n\kappa }$ as%
\begin{equation}
M^{2}-E_{n\kappa }^{2}+C_{ps}(M+E_{n\kappa })=-\kappa \left( \kappa
-1\right) \delta ^{2}d_{0}+\delta ^{2}\left( \frac{(M-E_{n\kappa
}+C_{ps})\Delta _{0}}{\delta ^{2}N_{1}}+\frac{N_{1}}{4}\right) ^{2},
\end{equation}%
where%
\begin{equation}
N_{1}=\left\{ 
\begin{array}{ccc}
2\left( n+\widetilde{l}+1\right) & \text{for} & \kappa >0 \\ 
2\left( n-\widetilde{l}\right) & \text{for} & \kappa <0%
\end{array}%
\right. ,\text{ }n=0,1,2,3,\cdots .
\end{equation}%
Thus, the energy spectrum can be obtained from the following energy
eigenvalue equation:%
\begin{equation}
\left[ 1+\left( \frac{\Delta _{0}}{N_{1}\delta }\right) ^{2}\right]
E_{n\kappa }^{2}-\left[ C_{ps}+\frac{2\Delta _{0}U}{N_{1}}\right] E_{n\kappa
}+\delta ^{2}\left[ U^{2}-\frac{SM}{\Delta _{0}}-\frac{\kappa (\kappa -1)}{12%
}\right] =0,
\end{equation}%
where%
\begin{equation}
U=\left( \frac{S}{N_{1}}+\frac{N_{1}}{4}\right) ,\text{ }S=\frac{%
(C_{ps}+M)\Delta _{0}}{\delta ^{2}},\text{ }\kappa (\kappa -1)=\widetilde{l}%
\left( \widetilde{l}+1\right) .
\end{equation}%
The two energy solutions of the above quadratic equation are 
\begin{equation}
E_{n\kappa }^{\pm }=\frac{\delta ^{2}\left( N_{1}^{2}C_{ps}+2N_{1}\Delta
_{0}U\right) \pm \delta ^{2}\sqrt{\left( N_{1}^{2}C_{ps}+2N_{1}\Delta
_{0}U\right) ^{2}+4N_{1}^{2}(\Delta _{0}^{2}+N_{1}^{2}\delta ^{2})\left( 
\frac{SM}{\Delta _{0}}+\frac{\kappa (\kappa -1)}{12}-U^{2}\right) }}{%
2(\Delta _{0}^{2}+N_{1}^{2}\delta ^{2})}.
\end{equation}%
For a given value of $n$ and $\kappa $ (or $\widetilde{l}$), the above
equation provides two distinct positive and negative energy spectra related
with $E_{n\kappa }^{+}$ or $E_{n\kappa }^{-}$, respectively. One of the
distinct solutions is only valid to obtain the negative-energy bound states
in the limit of the pseudospin symmetry. In the presence of exact pseudospin
symmetry ($C_{ps}=0$)$,$ we finally obtain%
\begin{equation}
E_{n\kappa }^{\pm }=\frac{M+\frac{1}{4}N_{1}^{2}\delta \pm N_{1}\sqrt{\left(
N_{1}^{2}+1\right) M^{2}-\left( M+\frac{1}{4}N_{1}^{2}\delta \right) ^{2}+%
\frac{1}{12}(N_{1}^{2}+1)\kappa (\kappa -1)\delta ^{2}}}{N_{1}^{2}+1}.
\end{equation}%
In this regards, states with various $n$ and $\widetilde{l}$ quantum numbers
having same energy spectrum are said to be degenerate states.

We calculate the negative bound state energy eigenvalues [4,35,36] from Eq.
(38) for several values of the quantum numbers $n$ and $\kappa (\widetilde{l}%
)$ in the pseudospin symmetry limit. They are displayed in Table 1. The
results have been calculated by using the following choices of parameters: $%
M=5.0$ $fm^{-1},$ $\Delta _{0}=3.40$ $fm^{-1}$ and $C_{ps}=-4.90$ $fm^{-1}$
[4]. From Table 1, one can clearly see that the degeneracy between two
states in the pseudospin doublets, i.e., $ns_{1/2},(n-1)d_{3/2}$ for $%
\widetilde{l}=1$ ($l=0$), $np_{3/2},(n-1)f_{5/2}$ for $\widetilde{l}=2$ ($%
l=1 $), $nd_{5/2},~(n-1)g_{7/2}$ for $\widetilde{l}=3$ ($l=2$)$,$ and $%
nf_{7/2},~(n-1)h_{9/2}$ for $\widetilde{l}=4$ ($l=3$)$,$ etc. Our numerical
approximations using the new approximation scheme, Eq.(25), are compared
with the ones obtained using Hulth$\mathbf{{\acute{e}}}$n square
approximation (see Eq. (34) of Ref. [4]). It is worth noting that such
approximation schemes are usually used in literature as effective
approximations to deal with the pseudo-centrifugal kinetic energy term in
the case of $\widetilde{l}>0$ and small $r$. One can easily see how the
approximation of the energy states is sensitive and dependent on the
approximation scheme used. Note that we have introduced a small positive
shift, $\delta ^{2}\widetilde{l}(\widetilde{l}+1)/12,$ to the conventional
approximation scheme [34], i.e., $r^{-2}=\delta W+W^{2},$ in calculating the
bound states (real energy states and corresponding wave functions).

Now, the corresponding wave functions calculations begin by calculating the
weight function from relation A8 as%
\begin{equation}
\rho (s)=\frac{1}{\sigma (s)}\exp \left( \int \frac{\tau (s)}{\sigma (s)}%
ds\right) =s^{2\epsilon _{n\kappa }}\left( 1-s\right) ^{2\kappa -1},
\end{equation}%
and the first part of the wave function:%
\begin{equation}
\phi (s)=\exp \left( \int \frac{\pi (s)}{\sigma (s)}ds\right) =s^{\epsilon
_{n\kappa }}\left( 1-s\right) ^{\kappa }.
\end{equation}%
Further the second part of the wave function can be obtained from relation
as 
\begin{equation}
y_{n_{r}}(s)=c_{n\kappa }s^{-2\epsilon _{n\kappa }}\left( 1-s\right)
^{-\left( 2\kappa -1\right) }\frac{d^{n}}{ds^{n}}\left[ s^{n+2\epsilon
_{n\kappa }}\left( 1-s\right) ^{n+2\kappa -1}\right] \sim P_{n_{r}}^{\left(
2\epsilon _{n\kappa },2\kappa -1\right) }(1-2s),
\end{equation}%
where $c_{n\kappa }$ is the normalization constant and $P_{n}^{\left( \mu
,\nu \right) }(x)$ are the Jacobi polynomials defined for $\func{Re}$($\nu
)>-1$ and $\func{Re}$($\mu )>-1$ in the interval $x\in \left[ -1,+1\right]
.. $ Using $G_{n\kappa }(s)=\phi (s)y_{n}(s),$ the lower-spinor wave
function reads%
\begin{equation*}
G_{n\kappa }(r)=c_{n\kappa }\left( \exp (-\epsilon _{n\kappa }\delta
r)\right) \left( 1-\exp (-\delta r)\right) ^{\kappa }P_{n}^{\left( 2\epsilon
_{n\kappa },2\kappa -1\right) }(1-2\exp (-\delta r))
\end{equation*}%
\begin{equation*}
=c_{n\kappa }\frac{\left( 2\epsilon _{n\kappa }+1\right) _{n}}{n!}\left(
\exp (-\epsilon _{n\kappa }\delta r)\right) \left( 1-\exp (-\delta r)\right)
^{\kappa }
\end{equation*}%
\begin{equation}
\times 
\begin{array}{c}
_{2}F_{1}%
\end{array}%
\left( -n,n+2\left( \epsilon _{n\kappa }+\kappa \right) ;1+2\epsilon
_{n\kappa };\exp (-\delta r)\right) ,\text{ }\kappa >0
\end{equation}%
with%
\begin{equation}
\epsilon _{n\kappa }\delta =\sqrt{M^{2}-E_{n\kappa }^{2}+C_{ps}(E_{n\kappa
}+M)+\kappa \left( \kappa -1\right) \delta ^{2}d_{0}}>0.
\end{equation}%
where $%
\begin{array}{c}
_{2}F_{1}%
\end{array}%
\left( -n,n+2\left( \epsilon _{n\kappa }+\kappa \right) ;1+2\epsilon
_{n\kappa };\exp (-\delta r)\right) $ is the hypergeometric series
terminates for $n=0$ and thus converges for all values of real parameters $%
\omega _{1}>0$ and $\widetilde{l}>0.$ When $C_{ps}=0,$ then $\epsilon
_{n\kappa }\delta =\sqrt{M^{2}-E_{n\kappa }^{2}+\kappa \left( \kappa
-1\right) \delta ^{2}d_{0}}$ with the restriction $E_{n\kappa
}^{2}<M^{2}+\kappa \left( \kappa -1\right) \delta ^{2}d_{0}$ is required to
obtain bound state (real) solutions for both positive and negative solutions
of $E_{n_{r}\kappa }$ in Eq. (39)$.$ Making use of the recurrence relation
of hypergeometric function%
\begin{equation}
\frac{d}{ds}\left[ 
\begin{array}{c}
_{2}F_{1}%
\end{array}%
\left( a;b;c;s\right) \right] =\left( \frac{ab}{c}\right) 
\begin{array}{c}
_{2}F_{1}%
\end{array}%
\left( a+1;b+1;c+1;s\right) ,
\end{equation}%
we obtain the corresponding upper component $F_{n_{r}\kappa }(r)$ from Eq.
(19) as%
\begin{equation*}
F_{n\kappa }(r)=b_{n\kappa }\frac{\left( \exp (-\epsilon _{n\kappa }\delta
r)\right) \left( 1-\exp (-\delta r)\right) ^{\kappa }}{(M-E_{n\kappa
}+C_{ps})}\left[ \frac{\kappa \delta \exp (-\delta r)}{\left( 1-\exp
(-\delta r)\right) }-\epsilon _{n\kappa }\delta -\frac{\kappa }{r}\right]
\end{equation*}%
\begin{equation*}
\times 
\begin{array}{c}
_{2}F_{1}%
\end{array}%
\left( -n,n+2\left( \epsilon _{n\kappa }+\kappa \right) ;1+2\epsilon
_{n\kappa };\exp (-\delta r)\right)
\end{equation*}%
\begin{equation*}
+b_{n\kappa }\left[ \frac{n\delta \left[ n+2\left( \kappa +\epsilon
_{n\kappa }\right) \right] \left( \exp (-\delta r)\right) ^{\epsilon
_{n\kappa }+1}\left( 1-\exp (-\delta r)\right) ^{\kappa }}{\left(
1+2\epsilon _{n\kappa }\right) (M-E_{n\kappa }+C_{ps})}\right]
\end{equation*}%
\begin{equation}
\times 
\begin{array}{c}
_{2}F_{1}%
\end{array}%
\left( 1-n,n+2\left( \epsilon _{n\kappa }+\kappa +\frac{1}{2}\right)
;2\left( 1+\epsilon _{n\kappa }\right) ;\exp (-\delta r)\right) ,
\end{equation}%
where $b_{n\kappa }$ is the new normalization factor. Based on the exact
pseudospin symmetry (i.e., when $C_{ps}=0,$ $E_{n\kappa }\neq M$), there are
only bound negative-energy states, otherwise the upper spinor component $%
F_{n\kappa }(r)$ will diverge. The energy solutions obtained from Eq.(38)
for a given values of $n$ and $\kappa $ need to be negative so that $%
G_{n\kappa }(r)$ and $F_{n\kappa }(r)$ are defined for the bound states,
i.e., $\epsilon _{n\kappa }>0,$ $\kappa \geq 1.$

\subsection{Spin Symmetry Solution}

The spin symmetry arises from $S(\vec{r})\sim V(\vec{r})$ in which the
nucleon move [4]. Therefore, we take the sum potential equal to the Hulth$%
\mathbf{{\acute{e}}}$n potential:%
\begin{equation}
\Sigma (r)=-\Sigma _{0}\frac{e^{-\delta r}}{1-e^{-\delta r}},\text{ }\Sigma
_{0}=V_{0}=Ze^{2}\delta ,
\end{equation}%
and apply the approximation in Eq.(25) dealing with the spin-orbit
centrifugal term $\kappa (\kappa +1)/r^{2}$. The choice $\Sigma
(r)=2V(r)\rightarrow V(r)$ enables one to restore the non-relativistic
solution when appropriate choice of parameter transformations is being
adopted [35]. Thus, Eq.(22) can be rewritten as%
\begin{equation}
\left[ \frac{d^{2}}{dr^{2}}-\kappa (\kappa +1)\delta ^{2}\left( d_{0}+\frac{%
e^{-\delta r}}{(1-e^{-\delta r})^{2}}\right) +\delta ^{2}\left( \nu _{2}^{2}%
\frac{e^{-\delta r}}{1-e^{-\delta r}}-\omega _{2}^{2}\right) \right]
F_{n\kappa }(r)=0,
\end{equation}%
with%
\begin{equation}
\nu _{2}^{2}=\frac{(M+E_{n\kappa }-C_{s})\Sigma _{0}}{\delta ^{2}},\text{ }%
\omega _{2}^{2}=\frac{M^{2}-E_{n\kappa }^{2}-C_{s}(M-E_{n\kappa })}{\delta
^{2}}.
\end{equation}%
Defining the new variable and parameters, 
\begin{equation*}
s=e^{-\delta r}\in \lbrack 0,+1],\text{ }A_{2}=\omega _{2}^{2}+\nu
_{2}^{2}+\kappa \left( \kappa +1\right) d_{0},\text{ }B_{2}=2\omega
_{2}^{2}+\nu _{2}^{2}+\kappa \left( \kappa +1\right) \left( 2d_{0}-1\right) ,%
\text{ }
\end{equation*}%
\begin{equation}
\varepsilon _{n\kappa }^{2}=\omega _{2}^{2}+\kappa \left( \kappa +1\right)
d_{0},
\end{equation}%
recasts Eq.(48) as 
\begin{equation}
\left( \frac{d^{2}}{ds^{2}}+\frac{1-s}{s(1-s)}\frac{d}{ds}+\frac{%
-A_{2}s^{2}+B_{2}s-\varepsilon _{n\kappa }^{2}}{s^{2}(1-s)^{2}}\right)
F_{n\kappa }(s)=0,\text{ }F_{n\kappa }(1)=0\text{ and }F_{n\kappa
}(0)\rightarrow 0.
\end{equation}%
Following the previous procedures, we obtain 
\begin{equation}
\widetilde{\tau }(s)=1-s,\text{\ }\sigma (s)=s(1-s),\text{\ }\widetilde{%
\sigma }(s)=-A_{2}s^{2}+B_{2}s-\varepsilon _{n\kappa }^{2},
\end{equation}%
and 
\begin{equation}
\pi (s)=\varepsilon _{n\kappa }-\left( \kappa +1+\varepsilon _{n\kappa
}\right) s,\text{ }k=\nu _{2}^{2}-\kappa \left( \kappa +1\right) -\left(
2\kappa +1\right) \varepsilon _{n\kappa },
\end{equation}%
\begin{equation}
\tau (s)=1+2\varepsilon _{n\kappa }-2\left( \varepsilon _{n\kappa }+\kappa +%
\frac{3}{2}\right) s,
\end{equation}%
Also, the parameters $\lambdabar $ and $\lambdabar _{n}$ take the forms:%
\begin{equation}
\lambdabar _{n}=n^{2}+2n\left( \varepsilon _{n\kappa }+\kappa +1\right) 
\text{ and }\lambdabar =\nu _{2}^{2}-\left( \kappa +1\right) (2\varepsilon
_{n\kappa }+\kappa +1),
\end{equation}%
giving the energy eigenvalue equation:%
\begin{equation}
\left[ 1+\left( \frac{\Sigma _{0}}{N_{2}\delta }\right) ^{2}\right]
E_{n\kappa }^{2}-\left[ C_{s}+\frac{2\Sigma _{0}W}{N_{2}}\right] E_{n\kappa
}+\delta ^{2}\left[ W^{2}+\frac{TM}{\Sigma _{0}}-\frac{\kappa (\kappa +1)}{12%
}\right] =0,
\end{equation}%
where 
\begin{subequations}
\begin{equation}
W=\left( \frac{T}{N_{2}}+\frac{N_{2}}{4}\right) ,\text{ }T=\frac{%
(C_{s}-M)\Sigma _{0}}{\delta ^{2}},\text{ }\kappa (\kappa +1)=l\left(
l+1\right)
\end{equation}%
\begin{equation}
N_{2}=\left\{ 
\begin{array}{ccc}
2\left( n+l+1\right) & \text{for} & \kappa >0 \\ 
2\left( n-l\right) & \text{for} & \kappa <0%
\end{array}%
\right. ,\text{ }n=0,1,2,3,\cdots .
\end{equation}%
The two energy solutions of the quadratic equation (56) can be obtained as 
\end{subequations}
\begin{equation}
E_{n\kappa }^{\pm }=\frac{\delta ^{2}\left( N_{2}^{2}C_{s}+2N_{2}\Sigma
_{0}W\right) \pm \delta ^{2}\sqrt{\left( N_{2}^{2}C_{s}+2N_{2}\Sigma
_{0}W\right) ^{2}+4N_{2}^{2}\left( \Sigma _{0}^{2}+N_{2}^{2}\delta
^{2}\right) \left( \frac{\kappa (\kappa +1)}{12}-\frac{TM}{\Sigma _{0}}%
-W^{2}\right) }}{2(\Sigma _{0}^{2}+N_{2}^{2}\delta ^{2})}.
\end{equation}%
For a given value of $n$ and $\kappa $ ($l$), we obtain two distinct
positive and negative energy spectra related with $E_{n\kappa }^{+}$ or $%
E_{n\kappa }^{-}$, respectively. However, the positive-energy solution is
valid for the spin symmetry limit. In the presence of exact spin symmetry ($%
C_{s}=0$)$,$ we can simply obtain%
\begin{equation}
E_{n\kappa }^{\pm }=\frac{-M+\frac{1}{4}N_{2}^{2}\delta \pm N_{2}\sqrt{%
\left( N_{2}^{2}+1\right) M^{2}-\left( -M+\frac{1}{4}N_{2}^{2}\delta \right)
^{2}+\frac{1}{12}(N_{1}^{2}+1)\kappa (\kappa +1)\delta ^{2}}}{N_{2}^{2}+1}.
\end{equation}%
Using Eq. (58), we calculate a few positive energy levels for various values
of quantum numbers $n$ and $\kappa (l))$ in the spin symmetry limit. In
Table 2, we present some numerical values with the following choices of
parameters: $M=5.0$ $fm^{-1},$ $\Sigma _{0}=3.40$ $fm^{-1}$ and $C_{s}=4.90$ 
$fm^{-1}.$ From Table 2, one can clearly see that the degeneracy between two
states in the spin doublets, i.e., $\left( np_{1/2},np_{3/2}\right) $ for $%
l=1$, $\left( nd_{3/2},nd_{5/2}\right) $ for $l=2$, $(nf_{5/2},~nf_{7/2})$
for $l=3,$ and $(ng_{7/2},~ng_{9/2})$ for $l=4,$ etc. For example, $0p_{1/2}$
with $n=0$ and $\kappa =1$ $(l=1)$ is the partner of $0p_{3/2}$ with $n=0$
and $\kappa =-2$ $(l=1).$

Next, we turn into the wave functions calculations. The calculated weight
function:%
\begin{equation}
\rho (s)=s^{2\varepsilon _{n\kappa }}\left( 1-s\right) ^{2\kappa +1},
\end{equation}%
enables us to write down the second part of the wave function as 
\begin{equation}
y_{n_{r}}(s)=a_{n\kappa }s^{-2\varepsilon _{n\kappa }}\left( 1-s\right)
^{-\left( 2\kappa +1\right) }\frac{d^{n}}{ds^{n}}\left[ s^{n+2\varepsilon
_{n\kappa }}\left( 1-s\right) ^{n+2\kappa +1}\right] \sim P_{n}^{\left(
2\varepsilon _{n\kappa },2\kappa +1\right) }(1-2s),
\end{equation}%
where $a_{n\kappa }$ is the normalization constant$.$ Furthermore, the first
part of the wave function reads 
\begin{equation}
\phi (s)=s^{\varepsilon _{n\kappa }}\left( 1-s\right) ^{\kappa +1}.
\end{equation}%
Thus, the upper component of the wave functions, $F_{n\kappa }(s)=\phi
(s)y_{n}(s),$ becomes%
\begin{equation*}
F_{n\kappa }(r)=a_{n\kappa }\left( \exp (-\varepsilon _{n\kappa }\delta
r)\right) \left( 1-\exp (-\delta r)\right) ^{\kappa +1}P_{n_{r}}^{\left(
2\varepsilon _{n\kappa },2\kappa +1\right) }(1-2\exp (-\delta r))
\end{equation*}%
\begin{equation}
=a_{n\kappa }\frac{\left( 2\varepsilon _{n\kappa }+1\right) _{n}}{n!}\left(
\exp (-\varepsilon _{n\kappa }\delta r)\right) \left( 1-\exp (-\delta
r)\right) ^{\kappa +1}\times 
\begin{array}{c}
_{2}F_{1}%
\end{array}%
\left( -n,n+2\left( \varepsilon _{n\kappa }+\kappa +1\right) ;1+2\varepsilon
_{n\kappa };\exp (-\delta r)\right) ,
\end{equation}%
where%
\begin{equation}
\varepsilon _{n\kappa }\delta =\sqrt{M^{2}-E_{n\kappa }^{2}+C_{s}(E_{n\kappa
}-M)+\kappa \left( \kappa +1\right) \delta ^{2}d_{0}}>0.
\end{equation}%
Note that the hypergeometric series $%
\begin{array}{c}
_{2}F_{1}%
\end{array}%
\left( -n,n+2\left( \varepsilon _{n\kappa }+\kappa +1\right) ;1+2\varepsilon
_{n\kappa };\exp (-\delta r)\right) $ is terminated for $n=0$ and thus it
converges for all values of real parameters $\omega _{2}>0$ and $\kappa >0.$
In case when $C_{s}=0,$ then $\varepsilon _{n\kappa }\delta =\sqrt{%
M^{2}-E_{n\kappa }^{2}+\kappa \left( \kappa +1\right) \delta ^{2}d_{0}}$
with a restriction for real bound states that $E_{n\kappa }^{2}<M^{2}+\kappa
\left( \kappa +1\right) \delta ^{2}d_{0}$ for both positive and negative
solutions of $E_{n\kappa }$ in Eq. (58)$.$ Thus, the corresponding
spin-symmetric lower-component $G_{n\kappa }(r)$ takes the form: 
\begin{equation*}
G_{n\kappa }(r)=b_{n\kappa }\frac{\left( \exp (-\varepsilon _{n\kappa
}\delta r)\right) \left( 1-\exp (-\delta r)\right) ^{\kappa +1}}{%
(M+E_{n\kappa }-C_{s})}\left[ \frac{\left( \kappa +1\right) \delta \exp
(-\delta r)}{\left( 1-\exp (-\delta r)\right) }-\varepsilon _{n\kappa
}\delta +\frac{\kappa }{r}\right]
\end{equation*}%
\begin{equation*}
\times 
\begin{array}{c}
_{2}F_{1}%
\end{array}%
\left( -n,n+2\left( \varepsilon _{n\kappa }+\kappa +1\right) ;1+2\varepsilon
_{n\kappa };\exp (-\delta r)\right)
\end{equation*}%
\begin{equation*}
+b_{n\kappa }\left[ \frac{n\delta \left[ n+2\left( \kappa +1+\varepsilon
_{n\kappa }\right) \right] \left( \exp (-\delta r)\right) ^{\varepsilon
_{n\kappa }+1}\left( 1-\exp (-\delta r)\right) ^{\kappa +1}}{\left(
1+2\varepsilon _{n\kappa }\right) (M+E_{n\kappa }-C_{s})}\right]
\end{equation*}%
\begin{equation}
\times 
\begin{array}{c}
_{2}F_{1}%
\end{array}%
\left( 1-n,n+2\left( \varepsilon _{n\kappa }+\kappa +\frac{3}{2}\right)
;2\left( 1+\varepsilon _{n\kappa }\right) ;\exp (-\delta r)\right) ,
\end{equation}%
where $E_{n\kappa }\neq -M$ when $C_{s}=0$, exact spin symmetry and $%
b_{n\kappa }$ is the normalization constant.

From the above expression, we see that there are only bound positive-energy
states, otherwise the lower spinor component $G_{n\kappa }(r)$ will diverge.
For a given values of $n$ and $\kappa ,$ we choose the suitable solution
that makes $G_{n\kappa }(r)$ and $F_{n\kappa }(r)$ satisfy the restriction
conditions for the bound states, i.e., $\epsilon _{n\kappa }>0,$ $\kappa
\geq -1$ and $E_{n\kappa }$ are positive.

A careful inspection of the relationship between the present set of
parameters $(\omega _{2}^{2},\nu _{2}^{2},A_{2},B_{2})$ and the previous set
of parameters $(\omega _{1}^{2},\nu _{1}^{2},A_{1},B_{1}).$ provides that
the spin symmetric positive energy solution can be simply obtained from the
pseudospin symmetric negative energy solution by making the replacements
[37]: 
\begin{equation*}
F_{n\kappa }(r)\leftrightarrow G_{n\kappa }(r),\text{ }V(r)\rightarrow -V(r)%
\text{ (or }\Sigma _{0}\leftrightarrow -\Delta _{0}\text{)},\text{ }\kappa
(\kappa +1)\leftrightarrow \kappa (\kappa -1)\text{ (or }\kappa
\leftrightarrow \kappa \pm 1\text{)},\text{ }
\end{equation*}%
\begin{equation}
C_{s}\leftrightarrow -C_{ps},\text{ }E_{n\kappa }^{+}\leftrightarrow
-E_{n\kappa }^{-},\text{ }\omega _{2}^{2}\leftrightarrow \omega _{1}^{2}%
\text{ and }\nu _{2}^{2}\leftrightarrow -\nu _{1}^{2}.
\end{equation}%
That is, with the above replacements, Eqs.(38) and (43) yield Eqs.(58) and
(63) and the vice versa is true.

Let us now present the non-relativistic limit. This can be achieved when we
set $C_{s}=0,$ $\kappa (\kappa +1)\rightarrow l(l+1),$ $\Sigma
_{0}=V_{0}=Ze^{2}\delta $ and using the mapping $E_{n\kappa }-M\simeq E_{nl}$
and \ $E_{n\kappa }+M\simeq 2m$ in Eqs.(59) and (63)$,$ then energy spectrum
(in atomic units $\hbar =c=e=1$) is%
\begin{equation}
E_{nl}=\frac{\delta ^{2}}{2m}\left\{ l\left( l+1\right) d_{0}-\left[ \frac{%
m\left( V_{0}/\delta ^{2}\right) }{\left( n+l+1\right) }-\frac{\left(
n+l+1\right) }{2}\right] ^{2}\right\} ,\text{ }n=0,1,2,\cdots \text{and }%
l=0,1,2,\cdots
\end{equation}%
where $n$ and $l$ are vibrational and orbital quantum numbers, respectively.
Also, the wave functions become%
\begin{equation*}
R_{nl}(r)=a_{nl}r^{-1}\left( \exp (-\sqrt{-2ME_{nl}+\frac{l\left( l+1\right)
\delta ^{2}}{12}}r)\right)
\end{equation*}%
\begin{equation*}
\times \left( 1-\exp (-\delta r)\right) ^{l+1}P_{n_{r}}^{\left( 2\sqrt{-%
\frac{2ME_{nl}}{\delta ^{2}}+\frac{l\left( l+1\right) }{12}},2l+1\right)
}(1-2\exp (-\delta r))
\end{equation*}%
\begin{equation*}
=a_{nl}\frac{\left( 2\sqrt{-\frac{2ME_{nl}}{\delta ^{2}}+\frac{l\left(
l+1\right) }{12}}+1\right) _{n}}{n!}r^{-1}\exp \left( -\sqrt{-\frac{2ME_{nl}%
}{\delta ^{2}}+\frac{l\left( l+1\right) }{12}}r\right) \left( 1-\exp
(-\delta r)\right) ^{l+1}
\end{equation*}%
\begin{equation}
\times 
\begin{array}{c}
_{2}F_{1}%
\end{array}%
\left( -n,n+2\left( \sqrt{-\frac{2ME_{nl}}{\delta ^{2}}+\frac{l\left(
l+1\right) }{12}}+l+1\right) ;1+2\sqrt{-\frac{2ME_{nl}}{\delta ^{2}}+\frac{%
l\left( l+1\right) }{12}};\exp (-\delta r)\right) ,
\end{equation}%
where $E_{nl}<l\left( l+1\right) \delta ^{2}/(24M).$ The traditional
approximation $\left( d_{0}=0\right) $ gives%
\begin{equation}
E_{nl}=-\frac{1}{2m}\left[ \frac{m}{\left( n+l+1\right) }-\frac{\left(
n+l+1\right) }{2}\delta \right] ^{2},
\end{equation}%
and%
\begin{equation*}
R_{nl}(r)=a_{nl}r^{-1}\left( \exp (-\varepsilon _{nl}r)\right) \left( 1-\exp
(-\delta r)\right) ^{l+1}P_{n_{r}}^{\left( 2\varepsilon _{nl}/\delta
,2l+1\right) }(1-2\exp (-\delta r))
\end{equation*}%
\begin{equation*}
=a_{nl}\frac{\left( 2\varepsilon _{nl}/\delta +1\right) _{n}}{n!}r^{-1}\exp
\left( -\varepsilon _{nl}r/\delta \right) \left( 1-\exp (-\delta r)\right)
^{l+1}
\end{equation*}%
\begin{equation}
\times 
\begin{array}{c}
_{2}F_{1}%
\end{array}%
\left( -n,n+2\left( \varepsilon _{nl}/\delta +l+1\right) ;1+2\varepsilon
_{nl}/\delta ;\exp (-\delta r)\right) ,
\end{equation}%
where $\varepsilon _{nl}=\sqrt{-2ME_{nl}},$ $E_{nl}<0$ for bound state
solution.

\section{Dirac Equation with Coupling to $r^{-1}$ Singular Orbital Term}

In the previous section we have found that the physical quantities like the
energy spectrum are critically dependent on the behavior of the system near
the singularity. That is why, for example, the energy spectrum depends
strongly on the angular momentum, which results from the $r^{-2}$
singularity of the orbital term, even for high excited states. Since the $%
r^{-2\text{ }}$ orbital term is too singular, then the validity of all such
approximations is limited only to very few of the lowest energy states.
Therefore, to extend accuracy to higher energy states one may attempt to
utilize the full advantage of the unique features of Dirac equation. For
example, the advantage of the Dirac equation over the Schr\"{o}dinger-like
equation is that the spin-orbit angular momentum singularity is $r^{-1}$
which is less singular than $r^{-2}.$ Therefore, it is more fruitful to
perform the analytic approximation of the orbital term in the Dirac equation
itself, which is a first-order differential equation, not in the resulting
second-order differential equation. The advantage is that in such case the
orbital term is less singular since it goes like $r^{-1}$ not like $r^{-2}.$
Therefore, one would expect that the solution of the Dirac equation is more
accurate by approximating the less singular distribution $r^{-1},$ which
makes it possible to extend the validity of the results to higher excitation
levels giving better analytic approximation for a wider energy spectrum [38].

Approximating the $r^{-1}$ spin-orbit term in the Dirac equation (Eqs.(18)
and (19)) by a function, say, $W(r)\approx r^{-1}$ results in the following
second order differential equations that should replace Eq. (22) and Eq.
(23) (in the relativistic units $\hbar =c=1$), respectively [38]%
\begin{equation}
\left( \frac{d^{2}}{dr^{2}}-\kappa ^{2}W^{2}(r)+\kappa \frac{dW(r)}{dr}%
-2\left( E_{n\kappa }+M_{s}\right) V(r)+E_{n\kappa }^{2}-M_{s}^{2}\right)
F_{n\kappa }(r)=0,
\end{equation}%
and%
\begin{equation}
\left( \frac{d^{2}}{dr^{2}}-\kappa ^{2}W^{2}(r)-\kappa \frac{dW(r)}{dr}%
-2\left( E_{n\kappa }-M_{ps}\right) V(r)+E_{n\kappa }^{2}-M_{ps}^{2}\right)
G_{n\kappa }(r)=0,
\end{equation}%
where $M_{s}=M-C_{s}$ and $M_{ps}=M+C_{ps}.$ Note that the resulting proper
approximation for the $r^{-2}$ term is not as trivial as one would think.
That is, the approximation for this term is not simply $W^{2}(r)$ but also
includes the derivative $dW(r)/dr$ giving the supersymmetric form $%
W^{2}(r)\pm W^{\prime }(r).$ To obtain an alternative solution using this
suggested proper approximation scheme, we may consider $W(r)=\delta /\left(
(e^{\delta r}-1\right) ,$ which is proportional to the Hulth$\mathbf{{\acute{%
e}}}$n potential. Therefore, we have applied the following proper
approximation introduced very recently by Alhaidari [38]%
\begin{equation}
\frac{\kappa \left( \kappa \pm 1\right) }{r^{2}}=\kappa ^{2}W^{2}(r)\mp
\kappa W^{\prime }(r).
\end{equation}

\subsection{Spin Symmetry solution}

We start by solving the Dirac-Hulth$\mathbf{{\acute{e}}}$n problem in the
presence of spin symmetry. We approximate the $r^{-1}$ orbital term by a
singular function $W(r)$ under certain approximation condition that will be
maintained throughout the subsection. If we define the variable $%
x=e^{-\delta r}\in \lbrack 0,+1]$ and inserting $V(r)=-V_{0}e^{-\delta
r}/(1-e^{-\delta r}),$ then the positive energy Schr\"{o}dinger-like
equation (71) in the new variable $x$ reads as follows%
\begin{equation}
\left( \frac{d^{2}}{dx^{2}}+\frac{1-x}{x(1-x)}\frac{d}{dx}+\frac{-\left(
\beta _{1}^{2}+\alpha _{1}^{2}+\kappa ^{2}\right) x^{2}+\left( \beta
_{1}^{2}+2\alpha _{1}^{2}-\kappa \right) x-\alpha _{1}^{2}}{x^{2}(1-x)^{2}}%
\right) F_{n\kappa }(x)=0,
\end{equation}%
provided that%
\begin{equation}
\alpha _{1}=\frac{1}{\delta }\sqrt{M_{s}^{2}-E_{n\kappa }^{2}},\text{ }\beta
_{1}=\frac{1}{\delta }\sqrt{2\left( E_{n\kappa }+M_{s}\right) V_{0}}.
\end{equation}%
Therefore, real solutions are possible only for $\left\vert E_{n\kappa
}\right\vert <M_{s}$ and potential strength $V_{0}>0$ (i.e., bound states).
Following the procedures explained in the previous section, we can find the
parametric constants of the NU as listed in Table 3. Further, the energy
equation can be obtained with the help of Table 3 and Ref. [22] as%
\begin{equation}
\sqrt{\left( M-C_{s}\right) ^{2}-E_{n\kappa }^{2}}=\frac{1}{N_{2}\delta }%
\left[ 2\left( E_{n\kappa }+M_{s}\right) V_{0}+\kappa ^{2}\delta ^{2}\right]
-\frac{N_{2}\delta }{4},
\end{equation}%
where $N_{2}$ is given in (57b). The above energy equation has the following
simple energy spectrum formula%
\begin{equation}
E_{n\kappa }^{\pm }=\frac{Q_{s}}{2P_{s}}\pm \sqrt{\frac{Q_{s}^{2}}{4P_{s}^{2}%
}+\frac{W_{s}}{P_{s}}},
\end{equation}%
with 
\begin{subequations}
\begin{equation}
Q_{s}=V_{0}\left[ \delta ^{2}(N_{s}^{2}-\kappa ^{2})-2V_{0}M_{s}\right] ,
\end{equation}%
\begin{equation}
P_{s}=V_{0}^{2}+\delta ^{2}N_{s}^{2},
\end{equation}%
\begin{equation}
W_{s}=M_{s}\left( M_{s}P_{s}+Q_{s}\right) +\frac{1}{4}\delta ^{4}\left[
\kappa ^{2}\left( 2N_{s}^{2}-\kappa ^{2}\right) -N_{s}^{4}\right] ,
\end{equation}%
\begin{equation}
N_{s}=\left\{ 
\begin{array}{ccc}
n+l+1 & \text{for} & \kappa >0 \\ 
n-l & \text{for} & \kappa <0%
\end{array}%
\right. ,\text{ }n=0,1,2,3,\cdots ,
\end{equation}%
where $Q_{s}^{2}+4P_{s}W_{s}\geq 0$ for real spectrum (bound states). For
numerical work, Eq. (77) and Eq. (78) are used to calculate a few positive
energy levels for various values of quantum numbers $n$ and $l$ in the spin
symmetry limit. We present some numerical values in Table 2 taking the
following values of parameters: $M=5.0$ $fm^{-1},$ $V_{0}=\Sigma _{0}=3.40$ $%
fm^{-1}$ and $C_{s}=4.90$ $fm^{-1}$ for the sake of comparison with the
previous $r^{-2}$ approximation results in Section IV$.$ In referring to
Table 2, it should be noted that the spectrum, in the $r^{-2}$ approximation
scheme, is wide with a fast transition toward the positive energy sector,
however, in the case of $r^{-1}$ approximation scheme, it is narrow with a
slow transition toward the positive energy.

Next, we calculate the upper component wave functions of Eq. (71) in the
form of hypergemetric function. Moreover, the nonrelativistic limit, is
obtained from Eq. (71) by setting $M_{s}\rightarrow M$ $(C_{s}=0)$, $\kappa
=l>0,$ $E_{n\kappa }+M\rightarrow 2m,$ $E_{n\kappa }-M\rightarrow E_{nl},$ $%
2V\rightarrow V.$ Therefore, Eq. (76) becomes 
\end{subequations}
\begin{equation}
E_{nl}=-\frac{\delta ^{2}}{2m}\left[ \frac{-m\left( V_{0}/\delta ^{2}\right)
+l^{2}}{\left( n+l+1\right) }-\frac{\left( n+l+1\right) }{2}\right] ^{2},
\end{equation}%
which is identical to Eq. (14) of Ref. [38] found for $V(r)=V_{0}/\left(
e^{\delta r}-1\right) .$ For the $S$-wave ($l=0$) restriction of (79)
reproduces the well-known nonrelativistic exact result [17]. The $l^{2}$
term is completely missing from the spectrum formula (67) because of the
approximation used for $r^{-2}$ and being substituted by another term $%
\delta ^{2}l(l+1)/(24m).$

Firstly, the weight function [33] reads 
\begin{equation}
\rho (x)=x^{2\alpha _{1}}(1-x)^{2\kappa +1},
\end{equation}%
which gives the first piece%
\begin{equation}
y_{n\kappa }(x)=P_{n}^{\left( 2\alpha _{1},2\kappa +1\right) }\left(
1-2x\right) =\frac{\Gamma (n+2\alpha _{1}+1)}{\Gamma (2\alpha _{1}+1)n!}%
\begin{array}{c}
_{2}F_{1}%
\end{array}%
\left( -n,n+2\left( \alpha _{1}+\kappa +1\right) ;1+2\alpha _{1};x\right) ,
\end{equation}%
and further the second piece reads%
\begin{equation}
\Phi (x)=x^{\alpha _{1}}\left( 1-x\right) ^{\gamma },\text{ }\gamma =\kappa
+1
\end{equation}%
where $\alpha _{2}$ and $\gamma $ are real positive parameters. Finally,
using Eq. (2), we can combine the two pieces as%
\begin{equation}
F_{n\kappa }(r)=A_{n\kappa }e^{-\delta \alpha _{1}r}\left( 1-e^{-\delta
r}\right) ^{\kappa +1}%
\begin{array}{c}
_{2}F_{1}%
\end{array}%
\left( -n,n+2\left( \alpha _{1}+\kappa +1\right) ;1+2\alpha _{1};e^{-\delta
r}\right) ,\text{ }n=0,1,2,\cdots .
\end{equation}%
where $A_{n\kappa }$ is the normalization factor and $\kappa =l$ for $\kappa
>0$ and $\kappa =-(l+1)$ for $\kappa <0$. The bound state solution requires
that the hypergeometric series terminate. The lower component wave functions
are calculated from Eq. (18) as%
\begin{equation*}
G_{n\kappa }(r)=\frac{A_{n\kappa }}{E_{n\kappa }+M_{s}}\left( \frac{\delta
(\kappa +1)e^{-\delta r}}{\left( 1-e^{-\delta r}\right) }-\delta \alpha _{1}+%
\frac{\kappa }{r}\right) F_{n\kappa }(r)
\end{equation*}%
\begin{equation*}
+A_{n\kappa }\frac{n\delta \left( n+2\alpha _{1}+2\kappa +2\right) }{\left(
E_{n\kappa }+M_{s}\right) \left( 1+2\alpha _{1}\right) }(1-e^{-\delta
r})^{\kappa +1}\left( e^{-\delta r}\right) ^{\alpha _{1}+1}
\end{equation*}%
\begin{equation}
\times 
\begin{array}{c}
_{2}F_{1}%
\end{array}%
\left( -n+1;n+2\left( \alpha _{1}+\kappa +1\right) +1;2\left( 1+\alpha
_{1}\right) ;e^{-\delta r}\right) .
\end{equation}%
Moreover, from Eq. (83), the nonrelativistic radial wave function reads%
\begin{equation}
R_{nl}(r)=A_{nl}e^{-\delta \lambda _{nl}r}\left( 1-e^{-\delta r}\right)
^{l+1}%
\begin{array}{c}
_{2}F_{1}%
\end{array}%
\left( -n,n+2\left( \lambda _{nl}+l+1\right) ;1+2\lambda _{nl};e^{-\delta
r}\right) ,\text{ }n=0,1,2,\cdots .
\end{equation}%
where $\lambda _{nl}=\sqrt{-2mE_{nl}}/\delta ,$ $E_{nl}<0$ which is defined
via (79).

\subsection{Pseudospin symmetry solution}

The pseudospin symmetry solutions could simply be found by applying the
following map on the spin symmetry solution in the previous section (both
the energy spectrum and the spinor wave functions) [35,38]%
\begin{equation}
F_{n\kappa }(r)\leftrightarrow G_{n\kappa }(r),\text{ }E_{n\kappa
}\rightarrow -E_{n\kappa },\text{ }\kappa \rightarrow -\kappa ,\text{ }%
V_{0}\rightarrow -V_{0},\text{ }C_{s}\leftrightarrow C_{ps},
\end{equation}%
which leads to the generation of Eq. (72) from Eq. (71). Making the change
of variables, $x=e^{-\delta r}\in \lbrack 0,+1],$ we can rewrite Eq. (72) as 
\begin{equation}
\left( \frac{d^{2}}{dx^{2}}+\frac{1-x}{x(1-x)}\frac{d}{dx}+\frac{-\left(
\alpha _{2}^{2}+\beta _{2}^{2}+\kappa ^{2}\right) x^{2}+\left( 2\alpha
_{2}^{2}+\beta _{2}^{2}+\kappa \right) x-\alpha _{2}^{2}}{x^{2}(1-x)^{2}}%
\right) G_{n\kappa }(x)=0,
\end{equation}%
provided that%
\begin{equation}
\alpha _{2}=\frac{1}{\delta }\sqrt{M_{ps}^{2}-E_{n\kappa }^{2}},\text{ }%
\beta _{2}^{2}=\frac{2\left( E_{n\kappa }-M_{ps}\right) V_{0}}{\delta ^{2}},%
\text{ }\left\vert E_{n\kappa }\right\vert <M_{s}.
\end{equation}%
Note that the parametric constants for the present case are listed in Table
3. The energy equation becomes%
\begin{equation}
\sqrt{\left( M+C_{ps}\right) ^{2}-E_{n\kappa }^{2}}=\frac{1}{N_{1}\delta }%
\left[ 2\left( E_{n\kappa }-M-C_{ps}\right) V_{0}+\kappa ^{2}\delta ^{2}%
\right] -\frac{N_{1}\delta }{4},
\end{equation}%
where $N_{1}$ is given in (35). The above energy equation has the following
simple energy spectrum formula%
\begin{equation}
E_{n\kappa }^{\pm }=\frac{Q_{ps}}{2P_{ps}}\pm \sqrt{\frac{Q_{ps}^{2}}{%
4P_{ps}^{2}}+\frac{W_{ps}}{P_{ps}}},
\end{equation}%
with 
\begin{subequations}
\begin{equation}
Q_{ps}=V_{0}\left[ \delta ^{2}(N_{ps}^{2}-\kappa ^{2})+2V_{0}M_{ps}\right] ,
\end{equation}%
\begin{equation}
P_{ps}=V_{0}^{2}+\delta ^{2}N_{ps}^{2},
\end{equation}%
\begin{equation}
W_{ps}=M_{ps}\left( M_{ps}P_{ps}-Q_{ps}\right) +\frac{1}{4}\delta ^{4}\left[
\kappa ^{2}\left( 2N_{ps}^{2}-\kappa ^{2}\right) -N_{ps}^{4}\right] ,
\end{equation}%
\begin{equation}
N_{ps}=\left\{ 
\begin{array}{ccc}
n+\widetilde{l}+1 & \text{for} & \kappa >0 \\ 
n-\widetilde{l} & \text{for} & \kappa <0%
\end{array}%
\right. ,\text{ }n=0,1,2,3,\cdots ,
\end{equation}%
where $Q_{ps}^{2}+4P_{ps}W_{ps}\geq 0$ for real spectrum (bound states).
Tables IV and V give approximation to the exact spin and pseudospin
symmetry, respectively. Further, we have provided two different
approximation models ( i.e., $r^{-2}$ and $r^{-1}$) for the sake of
comparison. The $r^{-2}$ approximation scheme (either conventional or
improved) is found to be more sensitive to spin-orbit quantum number $\kappa 
$ than the $r^{-1}$ proper approximation model [38]. The latter is found to
be valid for wide range energy spectrum (see Tables II, IV and V).%
\tablenotemark[1]%
\tablenotetext[1]{We have found a considerable discrepancy
in the numerical results of the two approximation schemes.}

The lower spinor can be found as 
\end{subequations}
\begin{equation}
G_{n\kappa }(r)=B_{n\kappa }e^{-\delta \alpha _{2}r}\left( 1-e^{-\delta
r}\right) ^{\kappa }%
\begin{array}{c}
_{2}F_{1}%
\end{array}%
\left( -n,n+2\left( \alpha _{2}+\kappa \right) ;1+2\alpha _{2};e^{-\delta
r}\right) ,\text{ }n=0,1,2,\cdots .
\end{equation}%
where $B_{n\kappa }$ is the normalization factor, $\kappa =\widetilde{l}+1$
for $\kappa >0$ and $\kappa =-\widetilde{l}$ for $\kappa <0$. The upper
spinor component wave functions are calculated from Eq. (19) as%
\begin{equation*}
G_{n\kappa }(r)=\frac{B_{n\kappa }}{M_{ps}-E_{n\kappa }}\left( \frac{\delta
\kappa e^{-\delta r}}{\left( 1-e^{-\delta r}\right) }-\delta \alpha _{2}-%
\frac{\kappa }{r}\right) G_{n\kappa }(r)
\end{equation*}%
\begin{equation*}
+B_{n\kappa }\frac{n\delta \left( n+2\alpha _{2}+2\kappa \right) }{\left(
M_{ps}-E_{n\kappa }\right) \left( 1+2\alpha _{2}\right) }(1-e^{-\delta
r})^{\kappa }\left( e^{-\delta r}\right) ^{\alpha _{2}+1}
\end{equation*}%
\begin{equation}
\times 
\begin{array}{c}
_{2}F_{1}%
\end{array}%
\left( -n+1;n+2\left( \alpha _{2}+\kappa \right) +1;2\left( 1+\alpha
_{2}\right) ;e^{-\delta r}\right) .
\end{equation}

\section{Results and Conclusions}

In the pseudospin symmetry case, the variation of the positive (negative)
energy spectrum $E_{n\widetilde{l}}^{+}$ ($E_{n\widetilde{l}}^{-}$) with the
screening parameter $\delta $ is shown in Fig. 2a (Fig. 2b), with a suitable
set of physical parameter values $C_{ps}=-4.90$ $fm^{-1},$ $M=5.0$ $fm^{-1}$
and $\Delta _{0}=3.40$ $fm^{-1}.$ For specific values of quantum numbers $n$
and $\widetilde{l},$ it is noted that when the screening parameter $\delta $
increases, the positive energy increases in the negative energy direction $%
\sim 0.03$ $fm^{-1},$ $\sim 0.13$ $fm^{-1}$ and $\sim 0.35$ $fm^{-1}$ for
pseudoorbital quantum numbers $\widetilde{l}=1,$ $\widetilde{l}=3$ and $%
\widetilde{l}=5,$ respectively, with a small energy difference between
states is small $(\sim 0.02-0.04$ $fm^{-1})$ when $\delta =0.20.$ However,
the negative energy spectrum increases in the negative energy direction $%
\sim 0.50$ $fm^{-1},$ $\sim 1.40$ $fm^{-1}$ and $\sim 1.90$ $fm^{-1}$ for $%
\widetilde{l}=1,$ $\widetilde{l}=3$ and $\widetilde{l}=5,$ respectively,
with energy difference $(\sim 0.45-0.50$ $fm^{-1})$ when $\delta =0.20.$ For
example, when $\delta =0.20,$ we have $E_{0\widetilde{l}}^{+}\sim 0.01,$ $%
0.03$ and $0.08$ $fm^{-1}$ and $E_{n\widetilde{l}}^{-}\sim 0.15,$ $0.60$ and 
$1.15$ $fm^{-1}$ with $\widetilde{l}=1,\widetilde{l}=3$ and $\widetilde{l}=5,
$ respectively. This large spacing returns to the new shifting energy term $%
\delta ^{2}\widetilde{l}(\widetilde{l}+1)/12.$ A more strongly binding
energy occurs for $E_{n\kappa }^{+}$ when $0<\delta <0.1$ (Fig.2a) but for $%
E_{n\kappa }^{-}$ when the screening parameter is lower, i.e., $0<\delta
<0.05$ (Fig. 2b)$.$

In the spin symmetry case, the variation of the energy spectra ($E_{n\kappa
}^{+}$ and $E_{n\kappa }^{-}$) with the screening parameter $\delta $ is
shown in Figs. 2a and b, with a suitable choice of physical parameter values 
$C_{s}=4.90$ $fm^{-1},$ $M=5.0$ $fm^{-1}$ and $\Sigma _{0}=3.40$ $fm^{-1}.$
The positive (negative) energy spectrum is plotted in Fig. 3a (Fig. 3b). For
specific values of quantum numbers $n$ and $\kappa (l),$ it is noted that
when the screening parameter $\delta $ increases, the positive energy
increases with a large amount in the positive energy direction $(\sim 4.6$ $%
fm^{-1})$ and difference between states is large $(\sim 1.3-2.0$ $fm^{-1})$
whereas the negative energy spectrum increases with small amount in the
positive energy direction $(\sim -0.10-0.05$ $fm^{-1})$ and difference in
energy spacing is $(\sim 0.55-0.85$ $fm^{-1}).$ For example, when $\delta
=0.20,$ we have $E_{n\kappa }^{+}\sim 0.2,$ $1.2$ and $2.4$ $fm^{-1}$ and $%
E_{n\kappa }^{-}\sim -0.095,$ $-0.085$ and $-0.065$ $fm^{-1}$ for the
orbital states $l=1,3$ and $5,$ respectively. These large (small) shifts
return to the new shifting energy term $\delta ^{2}l(l+1)/12.$ A more
strongly binding energy occurs for $E_{n\kappa }^{+}$ when $0<\delta <0.05$
(Fig. 3a) but for $E_{n\kappa }^{-}$ when the screening parameter is higher,
i.e., $0<\delta <0.1$ (Fig. 3b)$.$

Under the pseudospin symmetry, the energy-mass curves are plotted versus
mass for the pseudospin orbital quantum numbers $\widetilde{l}=1,$ $%
\widetilde{l}=3$ and$\ \widetilde{l}=5$ by taking the pseudospin constant $%
C_{ps}=-4.90$ $fm^{-1}$ and the screening parameter $\delta =0.25$ for a
given radial quantum number $n=0$ as shown in Fig. 4. There are two
different regions of energy spectrum ($E_{n\kappa }^{+}$ and $E_{n\kappa
}^{-})$ with the mass as shown in Fig. 4. In the positive energy part, $%
E_{n\kappa }^{+}$, nearly in the region $0<M<2.4$ $fm^{-1},$ the energy
spectrum is all in the negative region and the energy increases in the
direction of the positive energy as $\widetilde{l}$ increases. In the
negative energy part, $E_{n\kappa }^{-}$, nearly in the region $4.3<M<12$ $%
fm^{-1},$ the energy spectrum decreases in the direction of the negative
energy when $\widetilde{l}$ increases. Furthermore, under the spin symmetry,
the energy-mass curves are plotted versus mass for the orbital quantum
numbers $l=1,$ $l=3$ and$\ l=5$ by taking the spin constant $C_{s}=4.90$ $%
fm^{-1}$ and the screening parameter $\delta =0.25$ for a given radial
quantum number $n=0$ as shown in Fig. 5. There are two different regions of
energy spectrum ($E_{n\kappa }^{+}$ and $E_{n\kappa }^{-})$ versus mass as
shown in Fig. 5. In the negative energy part, $E_{n\kappa }^{-}$, nearly in
the region $0<M<0.6$ $fm^{-1},$ the energy spectrum is in the positive
region. The energy decreases in the direction of the negative energy as $l$
increases. In the positive energy part, $E_{n\kappa }^{+}$, nearly in the
region $3.0<M<12$ $fm^{-1},$ the energy spectrum increases in the direction
of the positive energy when $l$ increases.

In Fig. 6, we have plotted the energy spectrum versus the pseudospin
constant $C_{ps}$ for the parameters values $M=5.0$ $fm^{-1},$ $\Delta
_{0}=3.40$ $fm^{-1}$ and \ $\delta =0.25.$ The negative values of $C_{ps%
\text{ }}$show more strongly binding energies for $C_{ps\text{ }}<-10$ $%
fm^{-1}$ in the $E_{n\kappa }^{-}$ and less strongly binding energies for $%
C_{ps\text{ }}>-6$ $fm^{-1}$ in the $E_{n\kappa }^{+}$ for all $\widetilde{l}
$ values. The$\ $energy \ for the constants $-9$ $fm^{-1}<C_{ps\text{ }}<-5$ 
$fm^{-1}$ still show the negative energy up to the zero axis. Furthermore,
in Fig. 7, we have also plotted the energy spectrum versus the spin constant 
$C_{s}$ for the parameters values $M=5.0$ $fm^{-1},$ $\Sigma _{0}=3.40$ $%
fm^{-1}$ and \ $\delta =0.25.$ The positive values of $C_{s\text{ }}$show
more strongly binding energies for $C_{s\text{ }}>15$ $fm^{-1}$ in the $%
E_{n\kappa }^{-}$ and less strongly binding energies for $C_{s\text{ }}<10$ $%
fm^{-1}$ in the $E_{n\kappa }^{+}$ for all $l$ values. The$\ $energy \ for
the constants $-2$ $fm^{-1}<C_{s\text{ }}<-20$ $fm^{-1}$ still show the
negative energy up to the zero axis. For the case considered in Fig. 2a
(Fig. 2b) where $C_{ps}=-4.90$ $fm^{-1},$ $M=5.0$ $fm^{-1}$ and $\Delta
_{0}=3.40$ $fm^{-1}$ in the less strongly binding energies shows the
negative energy up to the zero axis and falls in the region $E_{n\kappa
}^{-}.$ However, the case where $C_{s}=4.90$ $fm^{-1},$ $M=5.0$ $fm^{-1}$
and $\Sigma _{0}=3.40$ $fm^{-1}$ considered in Fig. 3a (Fig. 3b) falls in
the less strongly binding energies in the region $E_{n\kappa }^{+}$.

We have seen that the Dirac equation for the Hulth$\mathbf{{\acute{e}}}$n
potential based on spin symmetry and pseudospin symmetry limitations can be
solved approximately for any arbitrary spin-orbital $\kappa $ state within
the framework of the Dirac theory. By using the basic ideas of the
parametric generalization of the NU method, the approximated positive and
negative energy eigenvalues for the arbitrary spin-orbital (pseudo-orbital)
angular momentum $l$ $\ (\widetilde{l}$ ) are obtained. \ An improved
approximation scheme is used to deal with the centrifugal $l(l+1)/r^{2}$
(pseudocentrifugal $\widetilde{l}(\widetilde{l}+1)/r^{2}$) potential term.
The energy spectrum for any $l$ ($\widetilde{l})$ states is obtained
analytically under the spin symmetry, $\Delta (r)=0$ (pseudospin symmetry, $%
\Sigma (r)=0$) limitations, the energy relations in the Dirac equation with
equal scalar and vector Hulth$\mathbf{{\acute{e}}}$n potentials are
recovered to see degenerate states. The relativistic bound state energy
eigenvalues and the correspondinf two-component spinor wave functions have
been easily reduced to the non-relativistic limits by applying appropriate
parameters replacements.

Finally, it is noted, from Tables 2, 4 and 5, that analytic solution of the
Dirac equation is more accurate by approximating the less singular
distribution spin-orbit angular momentum term $r^{-1},$ which makes it
possible to extend the validity of results to higher excitation levels
giving better analytic (numerical) approximation for a wider range spectrum
since the dependence of $r^{-1}$ on the angular quantum number is less than
the too singular term $r^{-2}.$ \newline
\acknowledgments The authors gratefully acknowledge TUBITAK for the partial
support. We thank the three referees for their enlightening suggestions
which greatly helped us to improve the paper.

\newpage

{\normalsize 
}

\bigskip

\bigskip

\bigskip \newpage

\bigskip {\normalsize 
}

\baselineskip= 2\baselineskip
\FRAME{ftbpFO}{0.0277in}{0.0277in}{0pt}{\Qct{A plot of the variation of the
singular orbital term $1/r^{2}$ (dotted-solid line) with (a) the
approximation of Ref. 4 (dash line), the conventional Greene-Aldrich of Ref.
36 (dash-dot line) and improved (solid line) approximations of $1/r^{2}$
with respect to $r$ where $\protect\delta =0.1$ $fm^{-1},$ (b) improved
approximation with various shifting constants.}}{}{Figure }{}\FRAME{ftbpFO}{%
0.0277in}{0.0277in}{0pt}{\Qct{The pseudospin symmetry energy spectrum versus
the screening parameter $\protect\delta $ for (a) $E_{n\protect\kappa }^{+}$
and (b) $E_{n\protect\kappa }^{-}$ in units of $fm^{-1}.$}}{}{Figure }{}%
\bigskip \FRAME{ftbpFO}{0.0277in}{0.0277in}{0pt}{\Qct{The spin symmetry
energy spectrum versus the screening parameter $\protect\delta $ for (a) $%
E_{n\protect\kappa }^{+}$ and (b) $E_{n\protect\kappa }^{-}$ in units of $%
fm^{-1}.$ }}{}{Figure }{}\bigskip \FRAME{ftbpFO}{0.0277in}{0.0277in}{0pt}{%
\Qct{The pseudospin symmetry energy spectrum versus the mass. The solid,
dash and dash-dot lines of the energy spectra including $E_{n\protect\kappa %
}^{+}$ and $E_{n\protect\kappa }^{-}$ are plotted for $\widetilde{l}=1,$ $%
\widetilde{l}=3$ and $\widetilde{l}=5,$ respectively$.$ }}{}{Figure }{}%
\bigskip \FRAME{ftbpFO}{0.0277in}{0.0277in}{0pt}{\Qct{The spin symmetry
energy spectrum versus the mass. The solid, dash and dash-dot lines energy
spectra including $E_{n\protect\kappa }^{+}$ and $E_{n\protect\kappa }^{-}$
are plotted for $l=1,$ $l=3$ and $l=5,$ respectively$.$ }}{}{Figure }{}%
\bigskip \FRAME{ftbpFO}{0.0277in}{0.0277in}{0pt}{\Qct{The pseudospin
symmetry energy spectrum versus the constant $C_{ps}.$ The solid, dash and
dash-dot lines energy spectra including $E_{n\protect\kappa }^{+}$ and $E_{n%
\protect\kappa }^{-}$ are plotted for $\widetilde{l}=1,$ $\widetilde{l}=3$
and $\widetilde{l}=5,$ respectively$.$ }}{}{Figure }{}\FRAME{ftbpFO}{0.0277in%
}{0.0277in}{0pt}{\Qct{The spin symmetry energy spectrum versus the constant $%
C_{s}.$ The solid, dash and dash-dot lines energy spectra including $E_{n%
\protect\kappa }^{+}$ and $E_{n\protect\kappa }^{-}$ are plotted for $l=1,$ $%
l=3$ and $l=5,$ respectively$.$ }}{}{Figure }{}\bigskip \newpage

\bigskip 
\begin{table}[tbp]
\caption{The negative bound state energy eigenvalues in units of $fm^{-1}$
of the pseudospin symmetry Hulth$\mathbf{{\acute{e}}}$n potential for
various values of $n,$ $\widetilde{l}=l+1$ and $\protect\delta .$ }%
\begin{tabular}{llllllllllll}
\tableline\tableline$\widetilde{l}$~ & ~$n,\kappa <0,\kappa >0$~ & ~$\delta $%
~~ & States & ~$E_{n\kappa }$ & $E_{n\kappa }$ [4] & $\widetilde{l}$~ & ~$%
n,\kappa <0,\kappa >0$ & ~$\delta $ & States & ~$E_{n\kappa }$ & $E_{n\kappa
}$ [4] \\ 
\tableline$1$ & $1,-1,2$ & $0.025$ & $(1s_{1/2},0d_{3/2})$ & $0.0972235$ & $%
0.0963638$ & $1$ & $2,-1,2$ & $0.025$ & $(2s_{1/2},1d_{3/2})$ & $0.0938034$
& $0.0928939$ \\ 
&  & $0.100$ &  & $0.0561798$ & $0.0425738$ &  &  & $0.100$ &  & $0.0038600$
& $-0.0103694$ \\ 
&  & $0.175$ &  & $-0.0302923$ & $-0.0710009$ &  &  & $0.175$ &  & $%
-0.1758970$ & $-0.2174930$ \\ 
&  & $0.250$ &  & $-0.1544010$ & $-0.2346580$ &  &  & $0.250$ &  & $%
-0.4125570$ & $-0.4920870$ \\ 
$2$ & $1,-2,3$ & $0.025$ & $(1p_{3/2},0f_{5/2})$ & $0.0937343$ & $0.0912282$
& $2$ & $2,-2,3$ & $0.025$ & $(2p_{3/2},1f_{5/2})$ & $0.0889591$ & $%
0.0863238 $ \\ 
&  & $0.100$ &  & $0.00275013$ & $-0.0363590$ &  &  & $0.100$ &  & $%
-0.0673920$ & $-0.1078600$ \\ 
&  & $0.175$ &  & $-0.1793260$ & $-0.2930130$ &  &  & $0.175$ &  & $%
-0.3590490$ & $-0.4732160$ \\ 
&  & $0.250$ &  & $-0.4196540$ & $-0.6351320$ &  &  & $0.250$ &  & $%
-0.7041020$ & $-0.9131390$ \\ 
$3$ & $1,-3,4$ & $0.025$ & $(1d_{5/2},0g_{7/2})$ & $0.0888560$ & $0.0839128$
& $3$ & $2,-3,4$ & $0.025$ & $(2d_{5/2},1g_{7/2})$ & $0.0827390$ & $%
0.0775818 $ \\ 
&  & $0.100$ &  & $-0.0690512$ & $-0.1447100$ &  &  & $0.100$ &  & $%
-0.1542610$ & $-0.2316110$ \\ 
&  & $0.175$ &  & $-0.3642070$ & $-0.5760950$ &  &  & $0.175$ &  & $%
-0.5611560$ & $-0.7705370$ \\ 
&  & $0.250$ &  & $-0.7148860$ & $-1.0984500$ &  &  & $0.250$ &  & $%
-0.9872420$ & $-1.3540100$ \\ 
$4$ & $1,-4,5$ & $0.025$ & $(1f_{7/2},0h_{9/2})$ & $0.08260190$ & $0.0744360$
& $4$ & $2,-4,5$ & $0.025$ & $(2f_{7/2},1h_{9/2})$ & $0.0751593$ & $%
0.0666955 $ \\ 
&  & $0.100$ &  & $-0.1564720$ & $-0.2784550$ &  &  & $0.100$ &  & $%
-0.2536460$ & $-0.3771030$ \\ 
&  & $0.175$ &  & $-0.5680850$ & $-0.8953110$ &  &  & $0.175$ &  & $%
-0.7673870$ & $-1.0870200$ \\ 
&  & $0.250$ &  & $-1.0019200$ & $-1.5671200$ &  &  & $0.250$ &  & $%
-1.2384300$ & $-1.7758200$ \\ 
\tableline &  &  &  &  &  &  &  &  &  &  & 
\end{tabular}%
\end{table}
\bigskip 
\begin{table}[tbp]
\caption{The positive bound state energy eigenvalues in units of $fm^{-1}$
of the spin-symmetry Hulth$\mathbf{{\acute{e}}}$n potential for various
values of $n,$ $l$ and $\protect\delta .${\protect\small \ }}%
\begin{tabular}{llllllllllll}
\tableline\tableline$l$~ & ~$n,\kappa <0,\kappa >0$~ & ~$\delta $~~ & States
& ~$E_{n\kappa }$ (present)\tablenotemark[1]\tablenotetext[1]{Improved
approximation to a more singular orbital term $r^{-2}$.} & $E_{n\kappa \text{
}}$ (present)\tablenotemark[2]\tablenotetext[2]{Proper approximation to a
less singular term $r^{-1}$ [38].} & $l$~ & ~$n,\kappa <0,\kappa >0$ & ~$%
\delta $ & States & ~$E_{n\kappa }$ (present)\tablenotemark[1] & $E_{n\kappa
}$ (present)\tablenotemark[2] \\ 
\tableline$1$ & $0,-2,1$ & $0.025$ & $(0p_{1/2},0p_{3/2})$ & $-0.0942003$ & $%
-0.0995915$ & $1$ & $1,-2,1$ & $0.025$ & $(1p_{1/2},1p_{3/2})$ & $-0.0869848$
& $-0.0989452$ \\ 
&  & $0.100$ &  & $-0.00840935$ & $-0.0935025$ &  &  & $0.100$ &  & $%
+0.1022580$ & $-0.0833617$ \\ 
&  & $0.175$ &  & $+0.1727090$ & $-0.0803626$ &  &  & $0.175$ &  & $%
+0.4825270$ & $-0.0506572$ \\ 
&  & $0.250$ &  & $+0.4336300$ & $-0.0607447$ &  &  & $0.250$ &  & $%
+0.9884020$ & $-0.00443345$ \\ 
$2$ & $0,-3,2$ & $0.025$ & $(0d_{3/2},0d_{5/2})$ & $-0.0869533$ & $%
-0.0984295 $ & $2$ & $1,-3,2$ & $0.025$ & $(1d_{3/2},1d_{5/2})$ & $%
-0.0768780 $ & $-0.0974023$ \\ 
&  & $0.100$ &  & $+0.1027630$ & $-0.0750704$ &  &  & $0.100$ &  & $%
+0.2514980$ & $-0.0590862$ \\ 
&  & $0.175$ &  & $+0.4840740$ & $-0.0249639$ &  &  & $0.175$ &  & $%
+0.8697760$ & $+0.0210900$ \\ 
&  & $0.250$ &  & $+0.9915680$ & $+0.0491605$ &  &  & $0.250$ &  & $%
+1.6152900$ & $+0.1346870$ \\ 
$3$ & $0,-4,3$ & $0.025$ & $(0f_{5/2},0f_{7/2})$ & $-0.0768308$ & $%
-0.0970491 $ & $3$ & $1,-4,3$ & $0.025$ & $(1f_{5/2},1f_{7/2})$ & $%
-0.0639221 $ & $-0.0956585$ \\ 
&  & $0.100$ &  & $+0.2522540$ & $-0.0534195$ &  &  & $0.100$ &  & $%
+0.4335670$ & $-0.0320936$ \\ 
&  & $0.175$ &  & $+0.8720970$ & $+0.0385481$ &  &  & $0.175$ &  & $%
+1.3001200$ & $+0.0980973$ \\ 
&  & $0.250$ &  & $+1.6200500$ & $+0.1706690$ &  &  & $0.250$ &  & $%
+2.2370300$ & $+0.2762140$ \\ 
$4$ & $0,-5,4$ & $0.025$ & $(0g_{7/2},0g_{9/2})$ & $-0.0638592$ & $%
-0.0952974 $ & $4$ & $1,-5,4$ & $0.025$ & $(1g_{7/2},1g_{9/2})$ & $%
-0.04815070$ & $-0.0935402$ \\ 
&  & $0.100$ &  & $+0.4345750$ & $-0.0262998$ &  &  & $0.100$ &  & $%
+0.6422870$ & $+0.000171676$ \\ 
&  & $0.175$ &  & $+1.3032300$ & $+0.1159560$ &  &  & $0.175$ &  & $%
+1.7441400$ & $+0.1871360$ \\ 
&  & $0.250$ &  & $+2.2434000$ & $+0.3130470$ &  &  & $0.250$ &  & $%
+2.8076500$ & $+0.4324460$ \\ 
\tableline &  &  &  &  &  &  &  &  &  &  & 
\end{tabular}%
\end{table}

\bigskip

\begin{table}[tbp]
\caption{Specific values of the NU constants based on the spin and
pseudospin symmetric Dirac-Hulth$\mathbf{{\acute{e}}}$n problem considering
the recently introduced proper approximation to the less singularity $r^{-1}$
orbital term.}%
\begin{tabular}{lll}
\tableline\tableline Spin symmetry: &  & Pseudospin symmetry: \\ 
\tableline$c_{1}=1$ &  & $c_{1}=1$ \\ 
$c_{2}=1$ &  & $c_{2}=1$ \\ 
$c_{3}=1$ &  & $c_{3}=1$ \\ 
c$_{4}=1$ &  & c$_{4}=1$ \\ 
$c_{5}=0$ &  & $c_{5}=0$ \\ 
$c_{6}=-\frac{1}{2}$ &  & $c_{6}=-\frac{1}{2}$ \\ 
$c_{7}=\frac{1}{4}+\alpha _{1}^{2}+\beta _{1}^{2}+\kappa ^{2}$ &  & $c_{7}=%
\frac{1}{4}+\alpha _{2}^{2}+\beta _{2}^{2}+\kappa ^{2}$ \\ 
$c_{8}=-2\alpha _{1}^{2}-\beta _{1}^{2}+\kappa $ &  & $c_{8}=-2\alpha
_{2}^{2}-\beta _{2}^{2}-\kappa $ \\ 
$c_{9}=\alpha _{1}^{2}$ &  & $c_{9}=\alpha _{2}^{2}$ \\ 
$c_{10}=\frac{1}{4}\left( 2\kappa +1\right) ^{2}$ &  & $c_{10}=\frac{1}{4}%
\left( 2\kappa -1\right) ^{2}$ \\ 
$c_{11}=2\alpha _{1}$ &  & $c_{11}=2\alpha _{2}$ \\ 
$c_{12}=2\kappa +1$ &  & $c_{12}=2\kappa -1$ \\ 
$c_{13}=\alpha _{1}$ &  & $c_{13}=\alpha _{2}$ \\ 
$c_{14}=\kappa +1$ &  & $c_{14}=\kappa $ \\ 
$c_{15}=2\kappa +1$ &  & $c_{15}=2\kappa -1$ \\ 
$c_{16}=\kappa +1$ &  & $c_{16}=\kappa $ \\ 
$\xi _{1}=\alpha _{1}^{2}+\beta _{1}^{2}+\kappa ^{2}$ &  & $\xi _{1}=\alpha
_{2}^{2}+\beta _{2}^{2}+\kappa ^{2}$ \\ 
$\xi _{2}=2\alpha _{1}^{2}+\beta _{1}^{2}-\kappa $ &  & $\xi _{2}=2\alpha
_{2}^{2}+\beta _{2}^{2}+\kappa $ \\ 
$\xi _{3}=\alpha _{1}^{2}=\delta ^{-2}\left( M_{s}^{2}-E_{n\kappa
}^{2}\right) $ &  & $\xi _{3}=\alpha _{2}^{2}=\delta ^{-2}\left(
M_{ps}^{2}-E_{n\kappa }^{2}\right) $ \\ 
\tableline &  & 
\end{tabular}%
\end{table}

\bigskip

\begin{table}[tbp]
\caption{Approximation of the negative bound state energy eigenvalues based
on the exact pseudospin symmetry ($C_{ps}=0$) Hulth$\mathbf{{\acute{e}}}$n
potential for various values of $n,$ $\widetilde{l}=l+1$ and $\protect\delta %
.$ }%
\begin{tabular}{llllllllllll}
\tableline\tableline$\widetilde{l}$~ & ~$n,\kappa <0,\kappa >0$~ & ~$\delta $%
~~ & States & ~$E_{n\kappa }$ (present)\tablenotemark[1]%
\tablenotetext[1]{Improved approximation to a more singular orbital term
$r^{-2}$.} & $E_{n\kappa }$ (present)\tablenotemark[2]%
\tablenotetext[2]{Proper approximation to a less singular term $r^{-1}$
[38].} & $\widetilde{l}$~ & ~$n,\kappa <0,\kappa >0$ & ~$\delta $ & States & 
~$E_{n\kappa }$ (present)\tablenotemark[1] & $E_{n\kappa }$ (present)%
\tablenotemark[2] \\ 
\tableline$1$ & $1,-1,2$ & $0.025$ & $(1s_{1/2},0d_{3/2})$ & $4.98403$ & $%
4.99611$ & $1$ & $2,-1,2$ & $0.025$ & $(2s_{1/2},1d_{3/2})$ & $4.97167$ & $%
4.99376$ \\ 
&  & $0.100$ &  & $4.75186$ & $4.93821$ &  &  & $0.100$ &  & $4.56926$ & $%
4.90141$ \\ 
&  & $0.175$ &  & $4.28511$ & $4.81377$ &  &  & $0.175$ &  & $3.81106$ & $%
4.70660$ \\ 
&  & $0.250$ &  & $3.66359$ & $4.62906$ &  &  & $0.250$ &  & $2.89559$ & $%
4.426637$ \\ 
$2$ & $1,-2,3$ & $0.025$ & $(1p_{3/2},0f_{5/2})$ & $4.97165$ & $4.99270$ & $%
2 $ & $2,-2,3$ & $0.025$ & $(2p_{3/2},1f_{5/2})$ & $4.95580$ & $4.98965$ \\ 
&  & $0.100$ &  & $4.56885$ & $4.88469$ &  &  & $0.100$ &  & $4.34617$ & $%
4.83772$ \\ 
&  & $0.175$ &  & $3.80980$ & $4.65663$ &  &  & $0.175$ &  & $3.28315$ & $%
4.52424$ \\ 
&  & $0.250$ &  & $2.89301$ & $4.32792$ &  &  & $0.250$ &  & $2.13127$ & $%
4.08931$ \\ 
$3$ & $1,-3,4$ & $0.025$ & $(1d_{5/2},0g_{7/2})$ & $4.95577$ & $4.98821$ & $%
3 $ & $2,-3,4$ & $0.025$ & $(2d_{5/2},1g_{7/2})$ & $4.93649$ & $4.98446$ \\ 
&  & $0.100$ &  & $4.34556$ & $4.81515$ &  &  & $0.100$ &  & $4.09036$ & $%
4.75851$ \\ 
&  & $0.175$ &  & $3.28126$ & $4.45771$ &  &  & $0.175$ &  & $2.73792$ & $%
4.30443$ \\ 
&  & $0.250$ &  & $2.12740$ & $3.96084$ &  &  & $0.250$ &  & $1.42801$ & $%
3.70026$ \\ 
$4$ & $1,-4,5$ & $0.025$ & $(1f_{7/2},0h_{9/2})$ & $4.93644$ & $4.98265$ & $%
4 $ & $2,-4,5$ & $0.025$ & $(2f_{7/2},1h_{9/2})$ & $4.91377$ & $4.97820$ \\ 
&  & $0.100$ &  & $4.08954$ & $4.73030$ &  &  & $0.100$ &  & $3.80963$ & $%
4.66464$ \\ 
&  & $0.175$ &  & $2.73540$ & $4.22266$ &  &  & $0.175$ &  & $2.20283$ & $%
4.05329$ \\ 
&  & $0.250$ &  & $1.42282$ & $3.54589$ &  &  & $0.250$ &  & $0.81097$ & $%
3.27673$ \\ 
\tableline &  &  &  &  &  &  &  &  &  &  & 
\end{tabular}%
\end{table}

\bigskip

\begin{table}[tbp]
\caption{Approximation of the positive bound state energy eigenvalues based
on the exact spin symmetry ($C_{s}=0$) Hulth$\mathbf{{\acute{e}}}$n
potential for various values of $n,$ $l$ and $\protect\delta .$%
{\protect\small \ }}%
\begin{tabular}{llllllllllll}
\tableline\tableline$l$~ & ~$n,\kappa <0,\kappa >0$~ & ~$\delta $~~ & States
& ~$E_{n\kappa }$ (present)\tablenotemark[1]\tablenotetext[1]{Improved
approximation to a more singular orbital term $r^{-2}$.} & $E_{n\kappa \text{
}}$ (present)\tablenotemark[2]\tablenotetext[2]{Proper approximation to a
less singular term $r^{-1}$ [38].} & $l$~ & ~$n,\kappa <0,\kappa >0$ & ~$%
\delta $ & States & ~$E_{n\kappa }$ (present)\tablenotemark[1] & $E_{n\kappa
}$ (present)\tablenotemark[2] \\ 
\tableline$1$ & $0,-2,1$ & $0.025$ & $(0p_{1/2},0p_{3/2})$ & $-4.98993$ & $%
-4.99731$ & $1$ & $1,-2,1$ & $0.025$ & $(1p_{1/2},1p_{3/2})$ & $-4.97738$ & $%
-4.99375$ \\ 
&  & $0.100$ &  & $-4.84099$ & $-4.95718$ &  &  & $0.100$ &  & $-4.64843$ & $%
-4.90078$ \\ 
&  & $0.175$ &  & $-4.52642$ & $-4.86979$ &  &  & $0.175$ &  & $-3.98679$ & $%
-4.70098$ \\ 
&  & $0.250$ &  & $-4.07294$ & $-4.73717$ &  &  & $0.250$ &  & $-3.10497$ & $%
-4.40441$ \\ 
$2$ & $0,-3,2$ & $0.025$ & $(0d_{3/2},0d_{5/2})$ & $-4.97737$ & $-4.99356$ & 
$2$ & $1,-3,2$ & $0.025$ & $(1d_{3/2},1d_{5/2})$ & $-4.95984$ & $-4.98857$
\\ 
&  & $0.100$ &  & $-4.64815$ & $-4.89773$ &  &  & $0.100$ &  & $-4.38924$ & $%
-4.81949$ \\ 
&  & $0.175$ &  & $-3.98590$ & $-4.69175$ &  &  & $0.175$ &  & $-3.31306$ & $%
-4.46248$ \\ 
&  & $0.250$ &  & $-3.10317$ & $-4.38588$ &  &  & $0.250$ &  & $-2.01110$ & $%
-3.94799$ \\ 
$3$ & $0,-4,3$ & $0.025$ & $(0f_{5/2},0f_{7/2})$ & $-4.95982$ & $-4.98847$ & 
$3$ & $1,-4,3$ & $0.025$ & $(1f_{5/2},1f_{7/2})$ & $-4.93736$ & $-4.98205$
\\ 
&  & $0.100$ &  & $-4.38880$ & $-4.81796$ &  &  & $0.100$ &  & $-4.07298$ & $%
-4.71859$ \\ 
&  & $0.175$ &  & $-3.31174$ & $-4.45782$ &  &  & $0.175$ &  & $-2.56340$ & $%
-4.17448$ \\ 
&  & $0.250$ &  & $-2.00840$ & $-3.93859$ &  &  & $0.250$ &  & $-0.92240$ & $%
-3.41830$ \\ 
$4$ & $0,-5,4$ & $0.025$ & $(0g_{7/2},0g_{9/2})$ & $-4.93733$ & $-4.98196$ & 
$4$ & $1,-5,4$ & $0.025$ & $(1g_{7/2},1g_{9/2})$ & $-4.91001$ & $-4.97411$
\\ 
&  & $0.100$ &  & $-4.07241$ & $-4.71713$ &  &  & $0.100$ &  & $-3.71030$ & $%
-4.59756$ \\ 
&  & $0.175$ &  & $-2.56164$ & $-4.17002$ &  &  & $0.175$ &  & $-1.78844$ & $%
-3.84019$ \\ 
&  & $0.250$ &  & $-0.91879$ & $-3.40926$ &  &  & $0.250$ &  & $+0.082117$ & 
$-2.83080$ \\ 
\tableline &  &  &  &  &  &  &  &  &  &  & 
\end{tabular}%
\end{table}

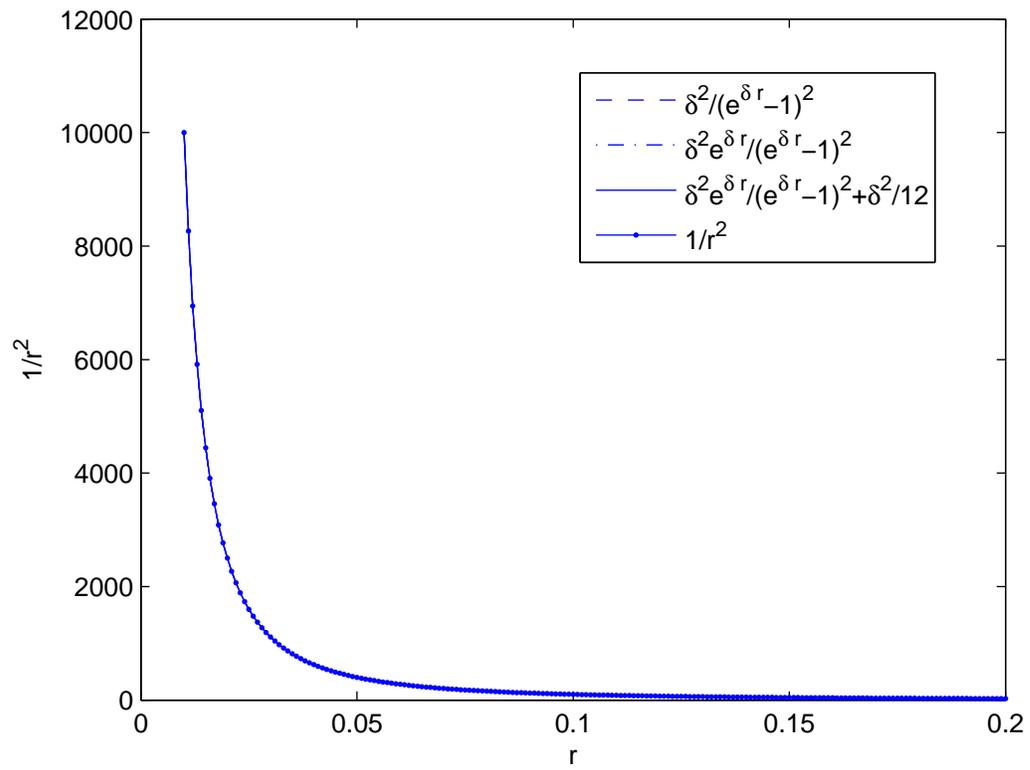
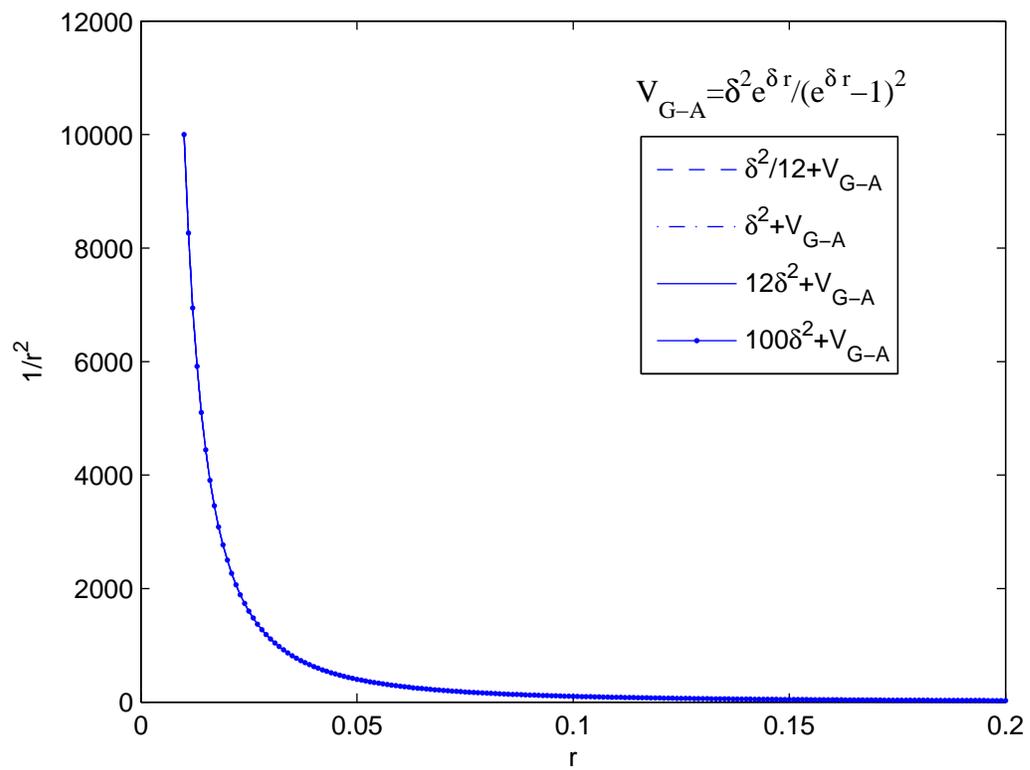
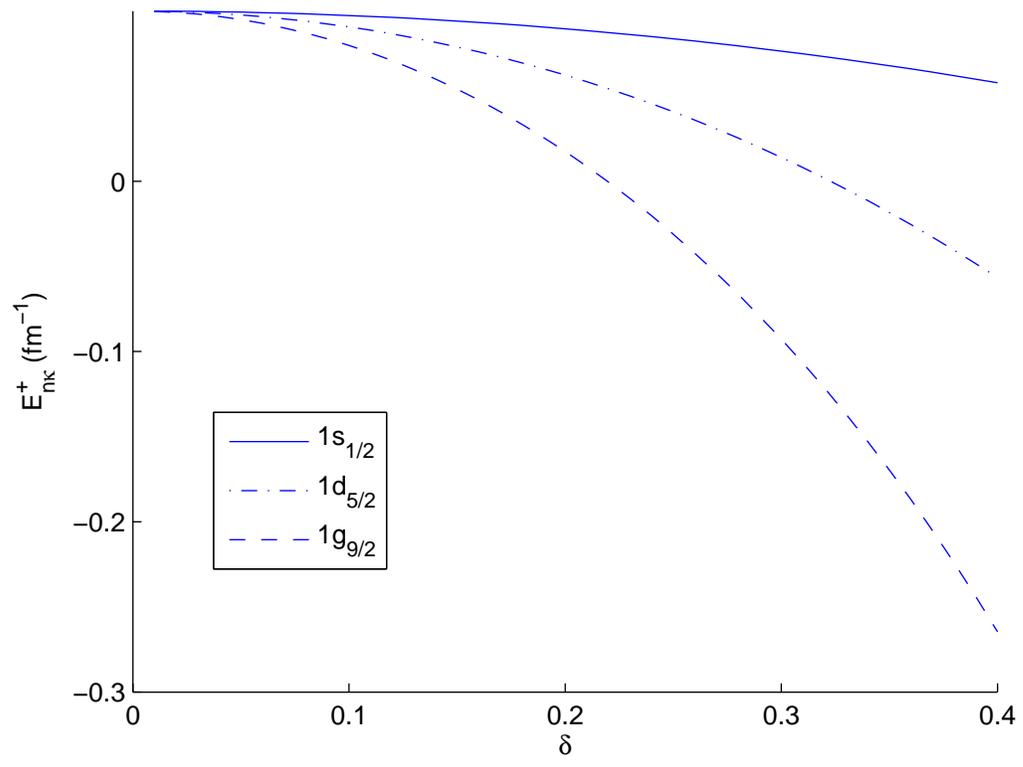
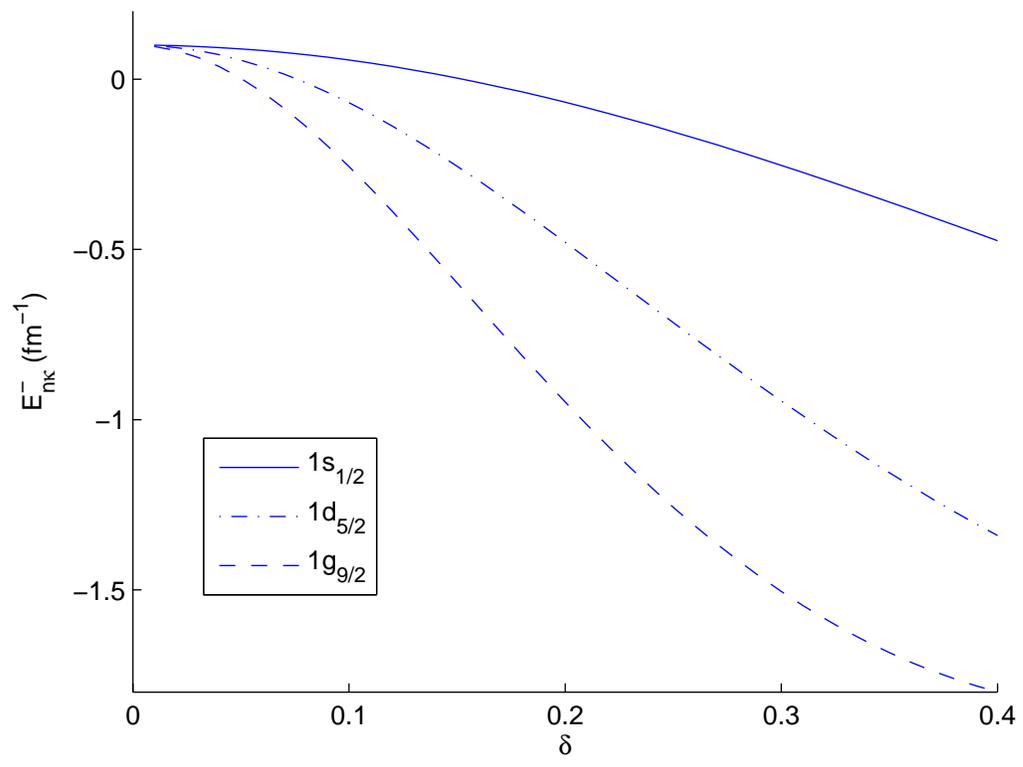
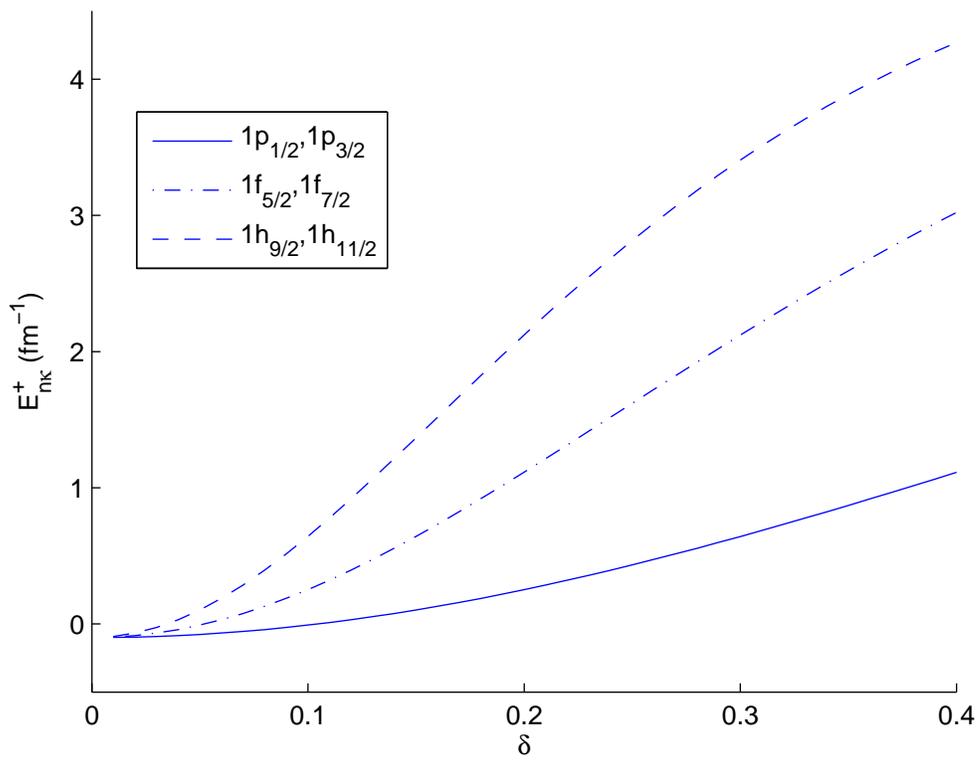
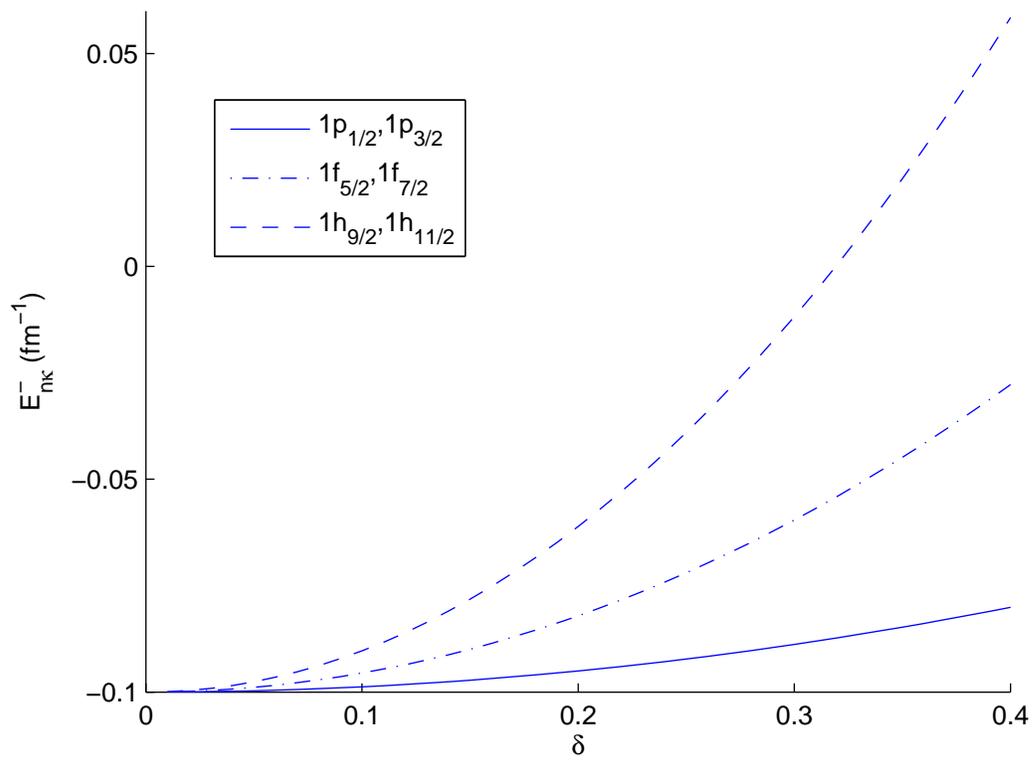
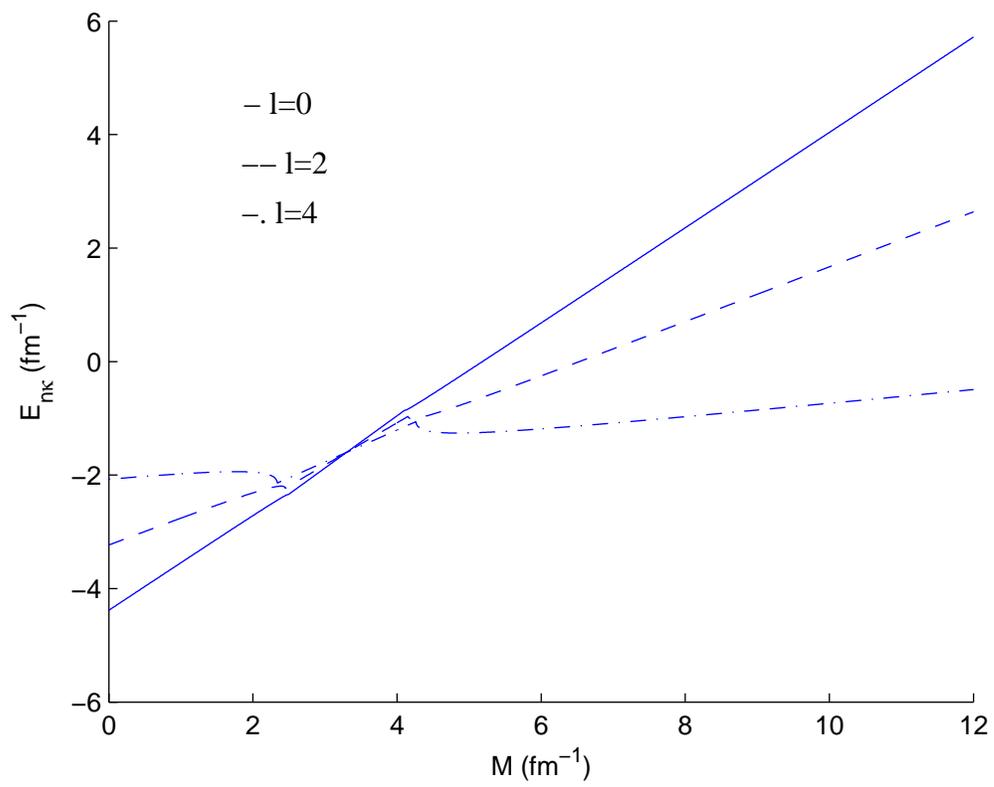
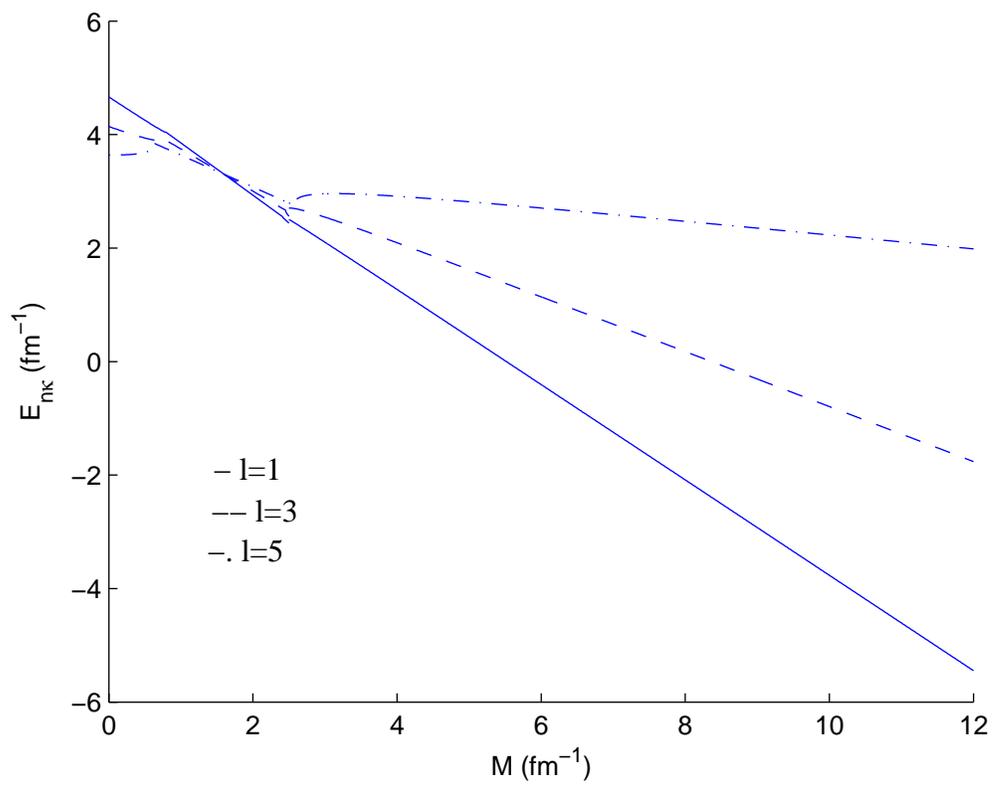
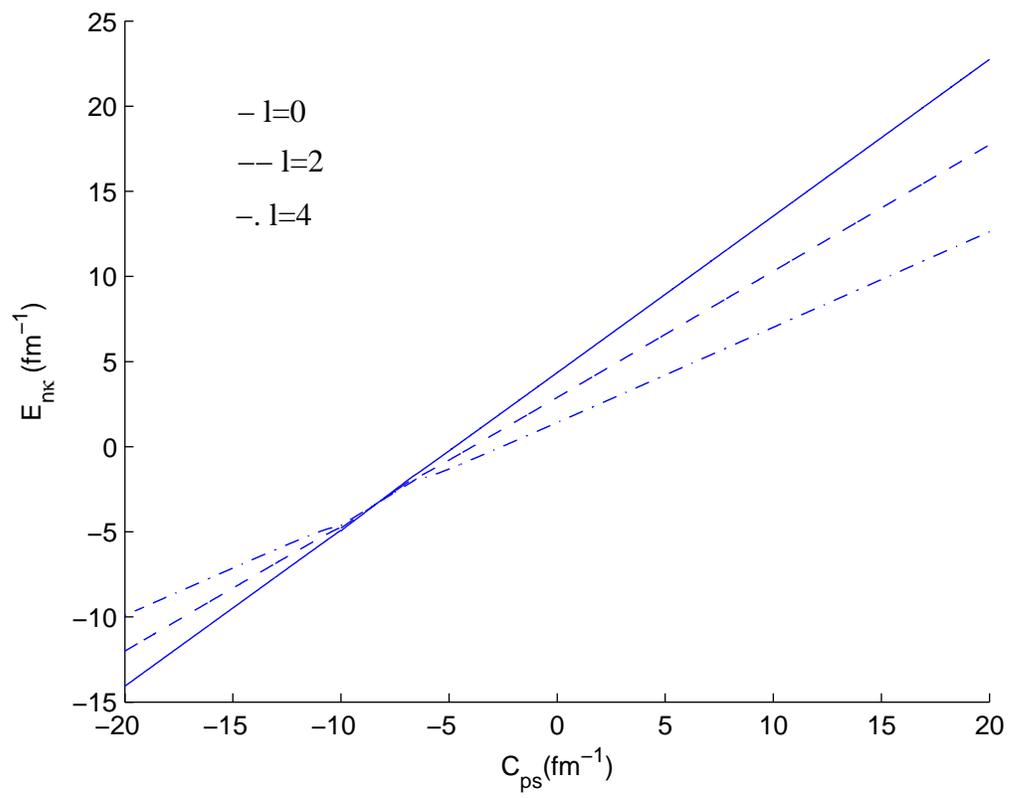
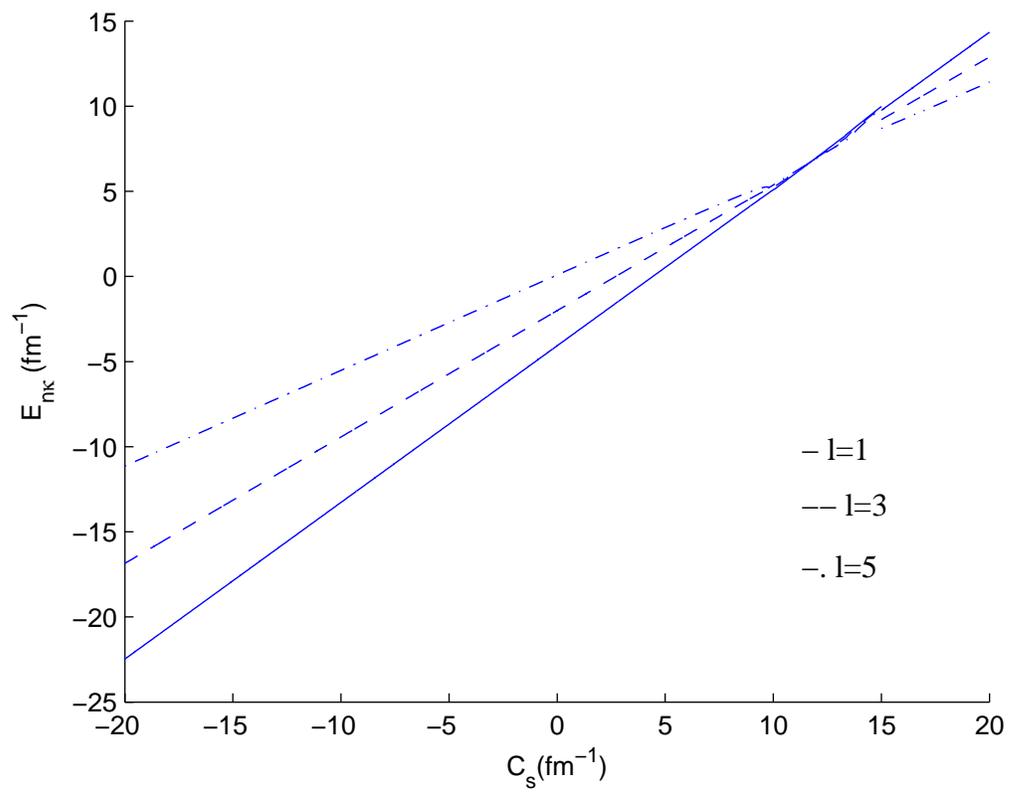
\end{document}